\documentclass{aa} 

\usepackage[utf8]{inputenc}
\usepackage{natbib}
\usepackage{graphicx}
\usepackage{color}
\usepackage{txfonts}
\usepackage{hyperref}
\usepackage{tabularx}
\usepackage{amsmath}
\usepackage{amssymb}
\usepackage{mathtools}
\usepackage[normalem]{ulem}
\usepackage{tikz}
\usetikzlibrary{arrows}
\usepackage{multirow}

\providecommand{\given}{\ensuremath{\hspace{0.07em}\mid\hspace{0.07em}}}
\providecommand{\parallax}{\ensuremath{\varpi}}
\providecommand{\parallaxsd}{\ensuremath{\sigma_\parallax}}
\providecommand{\parallaxtrue}{\ensuremath{\varpi_{\rm True}}}
\providecommand{\dist}{\ensuremath{r}}
\providecommand{\distest}{\ensuremath{\dist_{\rm est}}}
\providecommand{\distlen}{\ensuremath{L}}
\providecommand{\mg}{\ensuremath{M_{\rm G}}}
\providecommand{\gmag}{\ensuremath{G}}
\providecommand{\ag}{\ensuremath{A_{\rm G}}}

\newcommand\vect[1]{\ensuremath{\boldsymbol{#1}}}

\newcommand\gaia{\textit{Gaia~}}
\newcommand\hip{\textsc{Hipparcos}}

\newcommand\gdrone{\gaia~DR1~}
\newcommand\gdrtwo{\gaia~DR2~}

\newcommand\secref[1]{Sect.~\ref{#1}}
\newcommand\secrefalt[1]{Section~\ref{#1}}
\newcommand\figref[1]{Fig.~\ref{#1}}
\newcommand\figsref[1]{Figs.~\ref{#1}}
\newcommand\figrefalt[1]{Figure~\ref{#1}}

\newcommand\equref[1]{Eq.~\eqref{#1}}
\newcommand\equrefalt[1]{Equation~\eqref{#1}}
\newcommand\tabref[1]{Table~\ref{#1}}

\newcommand\repourl{https://github.com/agabrown/astrometry-inference-tutorials}
\newcommand\toprepo{\url{\repourl/}}
\newcommand\subrepo[1]{\url{\repourl/tree/master/#1}}

\begin{document}

%----------------------------------------------------------------
% Title, authors and abstract
%----------------------------------------------------------------
%\title{Distances from parallaxes}
%\subtitle{On the proper use of Gaia data}

\title{Gaia Data Release 2: Using Gaia parallaxes}
\subtitle{}

\author{
        X.        ~Luri                      \inst{\ref{inst:UB}}\relax
\and    A.G.A.        ~Brown                     \inst{\ref{inst:Leiden}}\relax
\and    L.M.      ~Sarro                     \inst{\ref{inst:UNED}}\relax
\and    F.        ~Arenou                    \inst{\ref{inst:Meudon}}\relax\and         C.A.L.          ~Bailer-Jones              \inst{\ref{inst:Heidelberg}}\relax
\and    A.        ~Castro-Ginard             \inst{\ref{inst:UB}}\relax
\and    J.        ~de Bruijne                \inst{\ref{inst:ESA}}\relax
\and    T.        ~Prusti                    \inst{\ref{inst:ESA}}\relax
\and    C.        ~Babusiaux                 \inst{\ref{inst:IPAG},\ref{inst:Meudon}}\relax
\and  H.E.      ~Delgado                   \inst{\ref{inst:UNED}}\relax
}

\institute{
    Dept. F\'{\i}sica Qu\`antica i Astrof\'{\i}sica, Institut de Ci\`encies del Cosmos (ICCUB), Universitat de Barcelona (IEEC-UB), Mart\'{\i} i Franqu\`es 1, E08028 Barcelona, Spain\relax \label{inst:UB}
    \and GEPI, Observatoire de Paris, Universit{\'e} PSL, CNRS, 5 Place Jules Janssen, 92190 Meudon, France\relax \label{inst:Meudon}
    \and Max Planck Institute for Astronomy, K\"onigstuhl 17, 69117 Heidelberg, Germany \relax \label{inst:Heidelberg}
    \and Sterrewacht Leiden, Leiden University, P.O.\ Box 9513, 2300 RA, Leiden, The Netherlands\relax \label{inst:Leiden}
    \and Science Support Office, Directorate of Science, European Space Research and Technology
    Centre (ESA/ESTEC), Keplerlaan 1, 2201 AZ, Noordwijk, The Netherlands\relax\label{inst:ESA}
    \and Dpto. Inteligencia Artificial, UNED, Juan del Rosal, 16, 28040 Madrid \relax \label{inst:UNED}
    \and Univ. Grenoble Alpes, CNRS, IPAG, 38000 Grenoble, France\relax \label{inst:IPAG}
}

\date{Received date / Accepted date}

\abstract{
  % context (optional)
  The second {\gaia} data release (\gdrtwo) provides precise five-parameter
  astrometric data (positions, proper motions, and parallaxes) for an
  unprecedented number of sources (more than $1.3$ billion, mostly stars).
  This new wealth of data will enable the undertaking of statistical analysis of
  many astrophysical problems that were previously infeasible for lack of  reliable astrometry, and in particular because of the lack of parallaxes. However, 
  the use of this wealth of astrometric data comes with a specific challenge:
  how can the astrophysical parameters of
  interest be properly inferred from these data?         
}{
  % aims        
  The main  focus of this paper, but not the only focus, is the issue of the estimation
  of distances from parallaxes, possibly combined with other information. We
  start with a critical review of the methods traditionally used to obtain
  distances from parallaxes and their shortcomings. Then we provide guidelines
  on how to use parallaxes more efficiently to estimate distances by using
  Bayesian methods. In particular  we also show that negative parallaxes, or
  parallaxes with relatively large uncertainties still contain valuable
  information. Finally, we provide examples that show more generally how to use
  astrometric data for parameter estimation, including the combination of proper
  motions and parallaxes and the handling of covariances in the uncertainties.
        
}{
  % methods
  The paper contains examples based on simulated {\it Gaia} data to illustrate
  the problems and the solutions proposed. Furthermore, the developments and
  methods proposed in the paper are linked to a set of tutorials included in the
  {\gaia} archive documentation that provide practical examples and a good
  starting point for the application of the recommendations to actual problems.
  In all cases the source code for the analysis methods is provided.

}{
    % results
  Our main recommendation is to always treat the derivation of (astro-)physical
  parameters from astrometric data, in particular when parallaxes are involved,
  as an inference problem which should preferably be handled with a full
  Bayesian approach.

}{
{\gaia} will provide fundamental data for many fields of astronomy. Further data
releases will provide more data, and more precise data. Nevertheless, to fully use the potential  it will always be necessary to pay careful attention to the statistical treatment of parallaxes and proper motions. The purpose of this
paper is to help astronomers find the correct approach.  }

\keywords{astrometry -- parallaxes -- Methods: data analysis -- Methods: statistical -- catalogues } 
\maketitle

%----------------------------------------------------------------
% Introduction
%----------------------------------------------------------------
\section{Introduction}
  \label{sec:intro}

The {\it Gaia} Data Release 2 (\gdrtwo) \citet{DR2-DPACP-36}
provides precise positions, 
proper motions, and parallaxes for an unprecedented number of objects
(more than 1.3 billion). Like Hipparcos \citet{1997ESASP1200.....E} in its day, the availability 
of a large amount of new astrometric data, and in particular parallaxes,
opens the way to revisit old astrophysical problems and to tackle new ones.
In many cases this will involve the inference of astrophysical quantities from \gaia astrometry, a
task that is less trivial than it appears, especially when parallaxes are involved. 

The naive use of the simple approach of inverting the parallax to estimate 
a distance can provide an acceptable estimate in a limited number of cases, 
in particular when  a precise parallax for an individual object is used.  However, one of the important contributions of \gdrtwo will
be the possibility of working with large samples of objects, all of them with 
measured parallaxes. In these cases a proper statistical treatment of the 
parallaxes in order to derive distances,  especially (but not only) when the 
relative uncertainties are large, is mandatory. Otherwise, the effects of the observational 
errors in the parallaxes can lead to potentially strong biases. More generally,
the use of full astrometric data to derive astrophysical parameters should
follow a similar approach. A proper statistical treatment of the data, its
uncertainties, and correlations is required to take full advantage of the \gaia results.

This paper is a complement for the {\it Gaia} consortium \gdrtwo papers.  We analyse the 
problem of the inference of distances (and other astrophysical parameters) from parallaxes.
In \secref{section:gaiadata} we start with a short review of the properties 
of the {\it Gaia} astrometric data. Then in \secref{sec:review} we review several of the most popular approaches to using measured parallaxes in astronomy
and highlight their intricacies, pitfalls, and problems. In \secref{sec:guide} we make
recommendations on what we think is the appropriate way to use astrometric
data. Finally, in \secref{sec:tutorial} we link to some worked examples, ranging 
from very basic demonstrations to full Bayesian analysis, available as Python and 
R notebooks and source code from the tutorial section on the {\it Gaia} archive
\footnote{\toprepo}.

%----------------------------------------------------------------
% TheData
%----------------------------------------------------------------
%----------------------------------------------------------------
% Describing the astrometric data
%----------------------------------------------------------------
\section{ {\it Gaia} astrometric data}
\label{section:gaiadata}

The {\it Gaia} astrometry,  i.e. celestial coordinates, trigonometric parallaxes, and proper motions for more than one billion objects,  results from the observations coming from the spacecraft instruments and their subsequent processing by the {\it Gaia} Data Processing and Analysis Consortium (DPAC). The astrometric processing is detailed in \citet{DR2-DPACP-51} and readers are strongly encouraged to familiarise themselves with the contents of that paper in order to understand the strengths and  weaknesses of the published astrometry, and in particular of the parallaxes. The processed data was submitted to extensive validation prior to publication, as detailed in \citet{DR2-DPACP-39}. This paper is also highly recommended  in order to gain a proper understanding of how to use and how {\rm not} to use the astrometric data. As a simple and striking example: a small number of sources with unrealistic very large positive and very large negative parallaxes are present in the data. Advice on how to filter these sources from the data analysis is provided in the \gdrtwo documentation.

%%%%%%%%%%%%%%%%%%%%%%%%%%%%%%%%%%%%%%%%%%%%%%%%%%%%
\subsection{Uncertainties}
\label{sec:uncertainities}

\begin{figure}[htb]
  \centering % trim = left bottom right top
  \includegraphics[width=1.00\columnwidth]{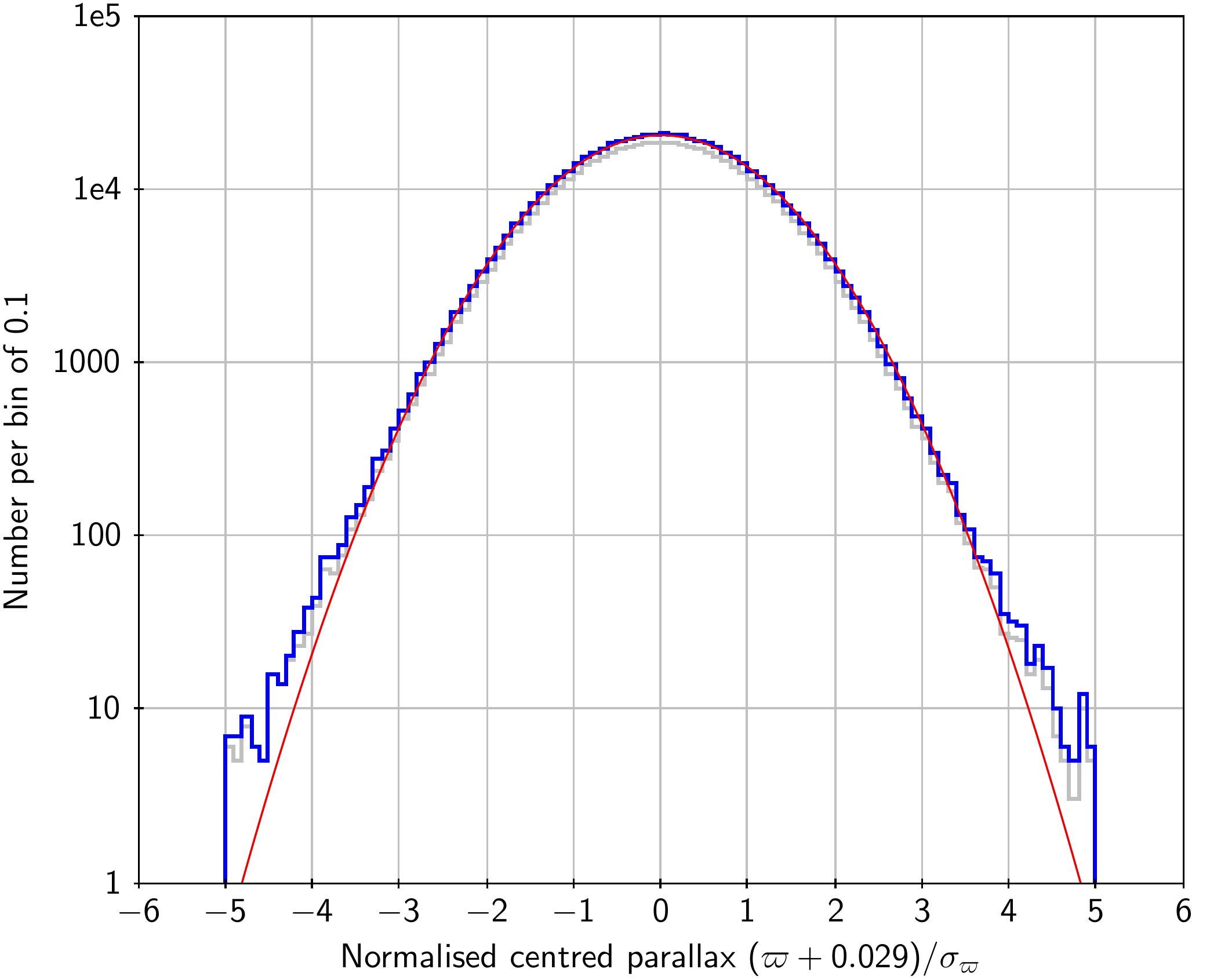}
  \caption{Distribution of normalised, re-centred parallaxes of $556\,849$ quasars from the AllWISE catalogue present in \gdrtwo (blue curve). The grey curve denotes the subsample composed of $492\,920$ sources with parallax errors $\sigma_\varpi < 1$~mas. The centring adopted in this plot reflects a global parallax zero-point shift of $-0.029$~mas. Ideally, both curves should follow a normal distribution with zero mean and unit variance. The red curve shows a Gaussian distribution with the same standard deviation (1.081) as the normalised centred parallaxes for the full sample. Figure from \citet{DR2-DPACP-51}.}
  \label{fig:QSO_Normalized_Parallaxes}
\end{figure}

The published parallaxes, and  more generally all astrometric parameters,  are measured quantities and as such have an associated measurement uncertainty. These uncertainties are published, source per source, and depend mostly on position on the sky as a result of the scanning law and on magnitude. For parallaxes, uncertainties are typically around 0.04~mas for sources brighter than $\sim$14~mag, around 0.1~mas for sources with a $G$ magnitude around 17, and around 0.7~mas at the faint end, around 20~mag. The astrometric uncertainties provided in \gdrtwo have been derived from the formal errors computed in the astrometric processing. Unlike for \gdrone, the parallax uncertainties have not been calibrated externally, i.e. they are known, as an ensemble, to be underestimated by $\sim$8--12\%\ for faint sources ($G \ga 16$~mag) outside the Galactic plane and by up to $\sim$30\%\  for bright stars ($G \la 12$~mag). Based on an assessment of the measured parallaxes of a set of about half a million known quasars, which can be assumed in practice to have zero parallax, the uncertainties are normally distributed with impressive approximation (\figref{fig:QSO_Normalized_Parallaxes}). However, as is common when taking measurements and especially in such large samples like the {\it Gaia} catalogue, there are small numbers of outliers, even up to unrealistically high confidence levels (e.g. at the 100$\sigma$ level).

%%%%%%%%%%%%%%%%%%%%%%%%%%%%%%%%%%%%%%%%%%%%%%%%%%%%
\subsection{Correlations}

The parallaxes for each source published in \gdrtwo have not been derived in isolation, but result from a simultaneous five-parameter fit of an astrometric source model to the data. In \gdrtwo, only one astrometric source model has been used, that of a single star. This model assumes a uniform, rectilinear space motion relative to the solar system barycentre. The astrometric data in \gdrtwo thus comprise five astrometric parameters\footnote{For a subset of the data, only two parameters (right ascension $\alpha$ and declination $\delta$) could be determined.} with their associated uncertainties, but also ten correlation coefficients between the estimated parameters. It is critical to use the full ($5 \times 5$) covariance matrix when propagating the uncertainties on subsets and/or linear combinations of the astrometric parameters.

As an example, consider the transformation of the measured proper motions $\mu_{\alpha*}$ and $\mu_\delta$ in equatorial coordinates to equivalent values $\mu_{l*}$ and $\mu_b$ in galactic coordinates. Following the notation in \citet[][Sections 1.2 and 1.5]{1997ESASP1200.....E}, we have
\begin{equation}
  \begin{pmatrix} 
    \mu_{l*} \\ \mu_b
  \end{pmatrix}
  =
  \begin{pmatrix} 
     c & s \\
    -s & c
  \end{pmatrix}
  \begin{pmatrix} 
    \mu_{\alpha*} \\ \mu_\delta
  \end{pmatrix},
\end{equation}
where the $2\times2$ matrix is a rotation matrix that depends on the object's coordinates $\alpha$ and $\delta$: $c = c(\alpha^\ast, \delta)$ and $s = s(\alpha^\ast, \delta)$. In order to transform the proper-motion errors from the equatorial to the galactic system, we have
\begin{eqnarray}
  \vec{C}_{l b} &=&
  \begin{pmatrix} 
\sigma_{\mu_{l *}}^2 & \rho_{\mu_{l *}}^{\mu_{b}} \sigma_{\mu_{l *}} \sigma_{\mu_b}\\
\rho_{\mu_{l *}}^{\mu_{b}} \sigma_{\mu_{l *}} \sigma_{\mu_b}     & \sigma_{\mu_b}^2
  \end{pmatrix}\\
&=&
  \vec{J}\vec{C}_{\alpha \delta}\vec{J}^\prime\\
&=&
  \begin{pmatrix} 
     c & s \\
    -s & c
  \end{pmatrix}
    \begin{pmatrix}
\sigma_{\mu_{\alpha *}}^2 & \rho_{\mu_{\alpha *}}^{\mu_{\delta}} \sigma_{\mu_{\alpha *}} \sigma_{\mu_\delta}\\
\rho_{\mu_{\alpha *}}^{\mu_{\delta}} \sigma_{\mu_{\alpha *}} \sigma_{\mu_\delta}     & \sigma_{\mu_\delta}^2
    \end{pmatrix}
    \begin{pmatrix} 
     c & -s \\
     s & c
  \end{pmatrix},
\end{eqnarray}
where the prime denotes matrix transposition, $\vec{J}$ denotes the Jacobian matrix of the transformation (which for a rotation is the rotation matrix itself), and $\vec{C}$ denotes the variance-covariance matrix. It immediately follows that $\sigma_{\mu_{l *}}$ and $\sigma_{\mu_{b}}$ depend on the generally non-zero correlation coefficient $\rho_{\mu_{\alpha *}}^{\mu_{\delta}}$ between the equatorial proper-motion measurements. Neglecting this correlation term can give seriously incorrect results.  Some further examples of how error propagation should be handled can be found in, for instance, \cite{1997ESASP.402...63B} and \cite{2000A&A...356.1119L}. In addition to error propagation, the covariance matrix should also be taken into account when estimating model parameters, for example in chi-square fitting, maximum likelihood estimates, Bayesian analysis, etc. For more details, see Volume 1, Section 1.5 of \citet{1997ESASP1200.....E}.

%As an example, consider the tangential velocity:
%\begin{equation}
%  v_{\rm T} = 4.74~ \mu~ / \parallax.
%\end{equation}
%
%The uncertainty on $v_{\rm T}$ not only depends on the uncertainties $\sigma_{\mu}$ on the proper motion $\mu$ and $\sigma_{\parallax}$ on the parallax \parallax\ but also on the correlation coefficient $\rho(\parallax, \mu)$. Using an approximation valid for small relative parallax errors this can be expressed as:
%
%\begin{equation}
%  \left(\frac{\sigma_{v_{\rm T}}}{v_{\rm T}}\right)^2 = \left(\frac{\sigma_{\mu}}{\mu}\right)^2 + \left(\frac{\sigma_{\parallax}}{\parallax}\right)^2 - 2~ \rho(\parallax,\mu) \left(\frac{\sigma_{\mu}}{\mu}\right) \left(\frac{\sigma_{\parallax}}{\parallax}\right).
%\end{equation}

%%%%%%%%%%%%%%%%%%%%%%%%%%%%%%%%%%%%%%%%%%%%%%%%%%%%
\subsection{Systematic errors}
\label{sec:systematics}

Both the design of the spacecraft and the design and implementation of the data processing software and algorithms aim to prevent biases or systematic effects in the astrometry. Systematic errors at low levels nonetheless exist in \gdrtwo \citep[see][]{DR2-DPACP-39,DR2-DPACP-51}. Systematic effects are complicated and largely unknown functions of position on the sky, magnitude, and colour.  Although systematic effects are not dealt with in the remainder of this paper, it is important for users to be aware of their presence.

The parallaxes and proper motions in \gdrtwo may be affected by systematic errors. Although the precise magnitude and distribution of these errors is unknown, they are believed to be limited, on global scales, to $\pm$0.1~mas for parallaxes and $\pm$0.1~mas~yr$^{-1}$ for proper motions. There is a significant average parallax zero-point shift of about $-30$~$\mu$as in the sense {\it Gaia} minus external data. This shift has {\em not} been corrected for and is present in the published data. Significant spatial correlations between stars, up to 0.04~mas in parallax and 0.07~mas~yr$^{-1}$ in proper motion, exist on both small ($\la$1$^\circ$) and intermediate ($\la$20$^\circ$) angular scales. As a result, {\bf averaging parallaxes over small regions of the sky, for instance in an open cluster, in the Magellanic Clouds, or in the Galactic Centre, will {\em not} reduce the uncertainty on the mean below the $\sim$$0.1$~mas level}.

Unfortunately, there is no simple recipe to account for the systematic errors. The general advice is to proceed with the analysis of the \gdrtwo data using the uncertainties reported in the catalogue,  ideally while modelling systematic effects as part of the analysis,  and to keep the systematics in mind when interpreting the results.

%%%%%%%%%%%%%%%%%%%%%%%%%%%%%%%%%%%%%%%%%%%%%%%%%%%%
\subsection{Completeness}
\label{sec:completeness}
 
As argued in the next sections, a correct estimation requires full knowledge of the survey selection function. Conversely, neglecting the selection function can causes severe biases. Derivation of the selection function is far from trivial, yet estimates have been made for \gdrone (TGAS) by, for instance, \citet{2017MNRAS.472.3979S} and \citet{2017MNRAS.470.1360B}.

This paper does not intend to define the survey selection function. We merely limit ourselves to mentioning a number of features of the \gdrtwo data that should be properly reflected in the selection function. The \gdrtwo catalogue is essentially complete between $G \approx 12$ and $\sim$17~mag. Although the completeness at the bright end ($G$ in the range $\sim$3--7~mag) has improved compared to \gdrone, a fraction of bright stars in this range is still missing in \gdrtwo. Most stars brighter than $\sim$3~mag are missing. In addition, about one out of every five high-proper-motion stars ($\mu \ga 0.6$~arcsec~yr$^{-1}$) is missing. Although the  onboard detection threshold at the faint end is equivalent to $G = 20.7$~mag, onboard magnitude estimation errors allow {\it Gaia} to see fainter stars, although not at each transit. \gdrtwo hence extends well beyond $G=20$~mag. However, in dense areas on the sky (above  $\sim$$400\,000$~stars~deg$^{-2}$),  the effective magnitude limit of the survey can be as bright as $\sim$18 mag. The somewhat fuzzy faint-end limit depends on object density (and hence celestial position) in combination with the scan-law coverage underlying the 22 months of data of \gdrtwo and the filtering on data quality that has been applied prior to publication. This has resulted in some regions on the sky showing artificial source-density fluctuations, for instance reflecting the scan-law pattern. In small, selected regions, gaps are present in the source distribution. These are particularly noticeable near very bright stars. In terms of effective angular resolution, the resolution limit of \gdrtwo is $\sim$0.4~arcsec.

Given the properties of \gdrtwo summarised above, the interpretation of the data is far from straightforward. This is particularly true when accounting for the incompleteness in any sample drawn from the \gaia\ Archive. We therefore strongly encourage the users of the data to read the papers and documentation accompanying \gdrtwo and to carefully consider the warnings given therein before drawing any conclusions from the data.

%----------------------------------------------------------------
% Review
%----------------------------------------------------------------
%----------------------------------------------------------------
% Historical review of solutions
%----------------------------------------------------------------
\section{Critical review of the traditional use of parallaxes}
  \label{sec:review}
  
We start this section by briefly describing how parallaxes are measured and how the presence of
measurement noise leads to the occurrence of zero and negative observed parallaxes. In the rest of
the section we review several of the most popular approaches to using measured parallaxes
($\parallax$) to estimate distances and other astrophysical parameters. In doing so we will attempt
to highlight the intricacies, pitfalls, and problems of these `traditional' approaches.

%%%%%%%%%%%%%%%%%%%%%%%%%%%%%%%%%%%%%%%%%%
\subsection{Measurement of parallaxes}
\label{sec:measurement}

In simplified form, astrometric measurements (source positions, proper motions, and parallaxes) are
made by repeatedly determining the direction to a source on the sky and modelling the change of
direction to the source as a function of time as a combination of its motion through space (as
reflected in its proper motion and radial velocity) and the motion of the observing platform (earth,
\gaia, etc.) around the Sun (as reflected in the parallax of the source). As explained in more detail
in \cite{2016A&A...595A...4L} and \cite{2012A&A...538A..78L}, this basic model of the source motion
on the sky describes the time-dependent coordinate direction from the observer towards an object
outside the solar system as the unit vector
\begin{equation}
  \vect{u}(t) = \langle \vect{r} + (t_\mathrm{B}-t_\mathrm{ep})
  (\vect{p}\mu_{\alpha*} + \vect{q}\mu_\delta + \vect{r}\mu_r) - 
  \varpi\vect{b}_\mathrm{O}(t)/A_\mathrm{u} \rangle\,,
  \label{eq:sourcemodel}
\end{equation}
where $t$ is the time of observation and $t_\mathrm{ep}$ is a reference time,  both in units of Barycentric
Coordinate Time (TCB); \vect{p}, \vect{q}, and \vect{r} are unit vectors pointing in the
direction of increasing right ascension, increasing declination, and towards the position
$(\alpha,\delta)$ of the source, respectively; $t_\mathrm{B}$ is the time of observation corrected
for the R{\o}mer delay; $\vect{b}_\mathrm{O}(t)$ is the barycentric position of the observer at the
time of observation; $A_\mathrm{u}$ is the astronomical unit; and  $\langle\rangle$ denotes
normalisation. The components of proper motion along \vect{p} and \vect{q} are respectively
$\mu_{\alpha*}=\mu_\alpha\cos\delta$ and $\mu_\delta$,  $\varpi$ is the parallax, and
$\mu_r=v_r\varpi/A_\mathrm{u}$ is the `radial proper motion' which accounts for the fact that the
distance to the star changes as a consequence of its radial motion, which in turn affects the
proper motion and parallax. The effect of the radial proper motion is negligibly small in most cases and can be
ignored in the present discussion.

\begin{figure}[htb]
  \centering
  \includegraphics[width=1.00\columnwidth]{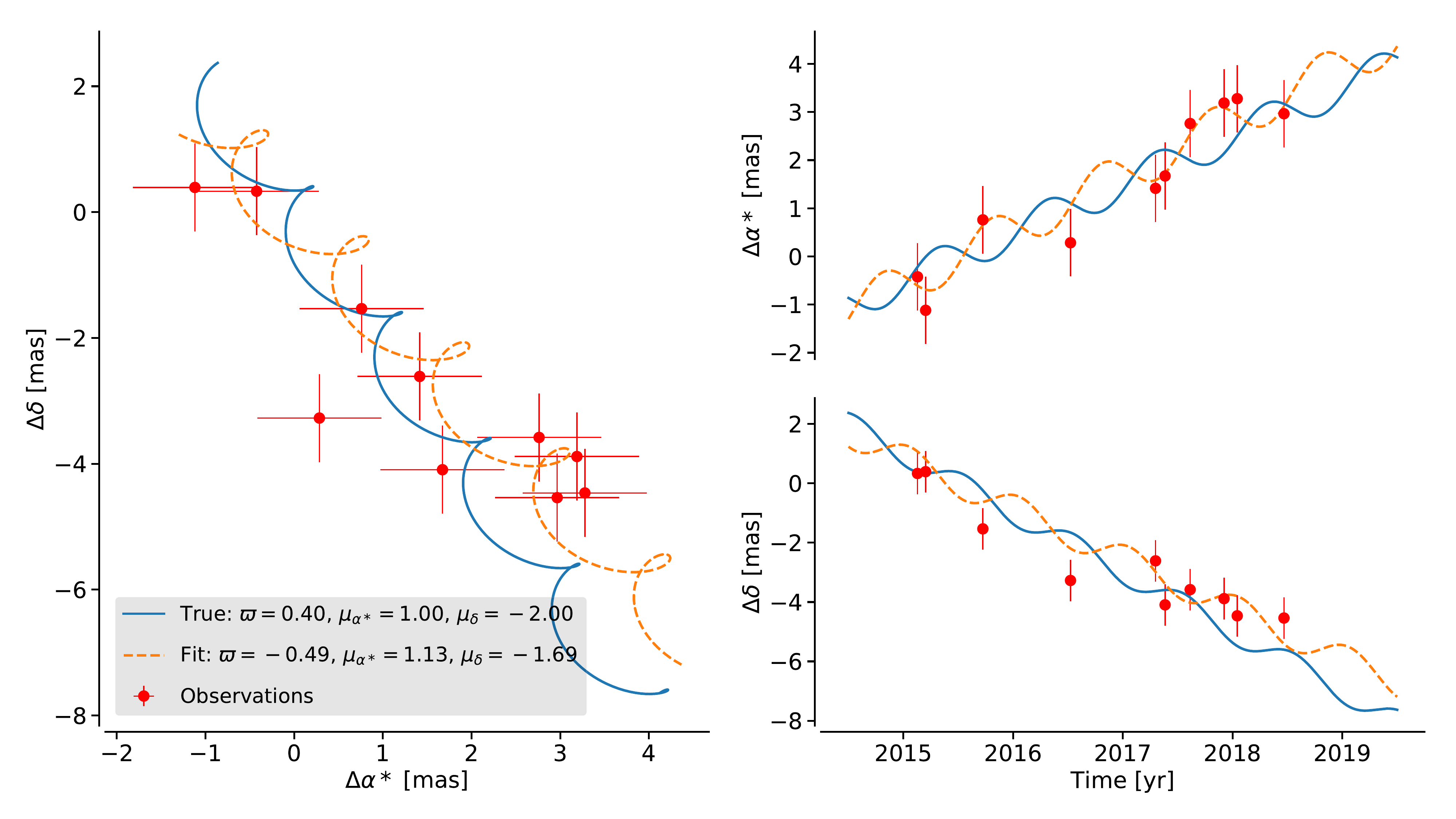}
  \caption{Example of a negative parallax arising from the astrometric data processing.  Solid blue
    lines, true path of the object; red dots, the individual measurements of the source position on
    the sky; dashed orange lines, the source path according to the least-squares astrometric
    solution, which here features a negative parallax. {\bf Left:} Path on the sky showing the effect
    of proper motion (linear trend) and parallax (loops). {\bf Right:} Right ascension and
    declination of the source as a function of time. In the fitted solution the
    negative parallax effect is equivalent to a yearly motion of the star in the opposite direction
  of the true parallactic motion (which gives a phase-shift of $\pi$ in the sinusoidal curves in the
  right panels). The error bars indicate a measurement uncertainty of $0.7$ mas, the
uncertainties on $\Delta\alpha*$ and $\Delta\delta$ are assumed to be uncorrelated.}
  \label{fig:path}
\end{figure}

The above source model predicts the well-known helix or wave-like pattern for the apparent motion
of a typical source on the sky. A fit of this model to noisy observations can lead to negative
parallaxes, as illustrated in \figref{fig:path}. We note how in the source model described in
\equref{eq:sourcemodel} the parallax appears as the factor $-\varpi$ in front of the barycentric
position of the observer, which means that for each source its parallactic motion on the sky will
have a sense which reflects the sense of the motion of the observer around the Sun. In the presence
of large measurement noise (comparable to the size of the parallax) it is entirely possible that the
parallax value estimated for the source model vanishes or becomes negative. This case can be
interpreted as the measurement being consistent with the source going `the wrong way around' on the
sky, as shown in \figref{fig:path}.

This example is intended to clarify why parallaxes can have non-positive observed values and, more
importantly, to convey the message that the parallax is not a direct measurement of the distance to
a source. The distance (or any other quantity depending on distance) has to be {\em estimated} from
the observed parallax (and other relevant information), taking into account the uncertainty in the
measurement. A simplified demonstration of how negative parallaxes arise (allowing the reader to reproduce
\figref{fig:path}) can be found in the online tutorials accompanying this paper
\footnote{\subrepo{luminosity-calibration/DemoNegativeParallax.ipynb}}.

%%%%%%%%%%%%%%%%%%%%%%%%%%%%%%%%%%%%%%%%%%
\subsection{Estimating distance by inverting the parallax}
\label{sec:parallaxinversion}

In the absence of measurement uncertainties, the distance to a star
can be obtained from its true parallax through $r=1/\parallaxtrue$, with
{\parallaxtrue} indicating the true value of the parallax. Thus,
naively we could say that the distance to a star can be obtained by
inverting the observed parallax, $\rho=1/\varpi$, where now $\rho$ is
used to indicate the distance derived from the observed value of the
parallax. For this discussion the observed parallax is assumed to be
free of systematic measurement errors and to be distributed normally
around the true parallax

\begin{equation}
   p(\parallax \mid \parallaxtrue) = \frac{1}{\parallaxsd \sqrt{2 \pi}}   \exp\left({-\frac{(\parallax - \parallaxtrue)^2}{2\parallaxsd^2}}\right)\,,
         \label{eq:pdf_truepi}
\end{equation}

\noindent where {\parallaxsd} indicates the measurement uncertainty on
{\parallax}. Blind use of $1/\parallax$ as an estimator of the
distance will  lead to unphysical results in case the
observed parallax is non-positive. Nevertheless, we could still
consider the use of the $1/\parallax$ distance estimate for positive
values, for instance, a sample where most or  all  of the
observed values are positive or, in the limiting case, where there is a single
positive parallax value. In this case, it is crucial to be aware
of the statistical properties of the estimate $\rho$. Given a true distance
$r=1/\parallaxtrue$, what will be the behaviour of $\rho$? We can
obtain the probability density function (PDF) of $\rho$ from
\equref{eq:pdf_truepi} as

\begin{eqnarray}
   p( \rho \mid \parallaxtrue ) & = & p(\parallax=1/\rho \mid \parallaxtrue)\cdot
   \left|\frac{d\parallax}{d\rho}\right| \nonumber \\ & = & \frac{1}{\rho^2
     \parallaxsd \sqrt{2 \pi}} \exp\left({-\frac{(1/\rho -
       \parallaxtrue)^2}{2\parallaxsd^2}}\right) 
         \label{eq:pdf_rho}
\end{eqnarray}

\begin{figure}[htb]
        \centering
        \includegraphics[width=1.00\columnwidth]{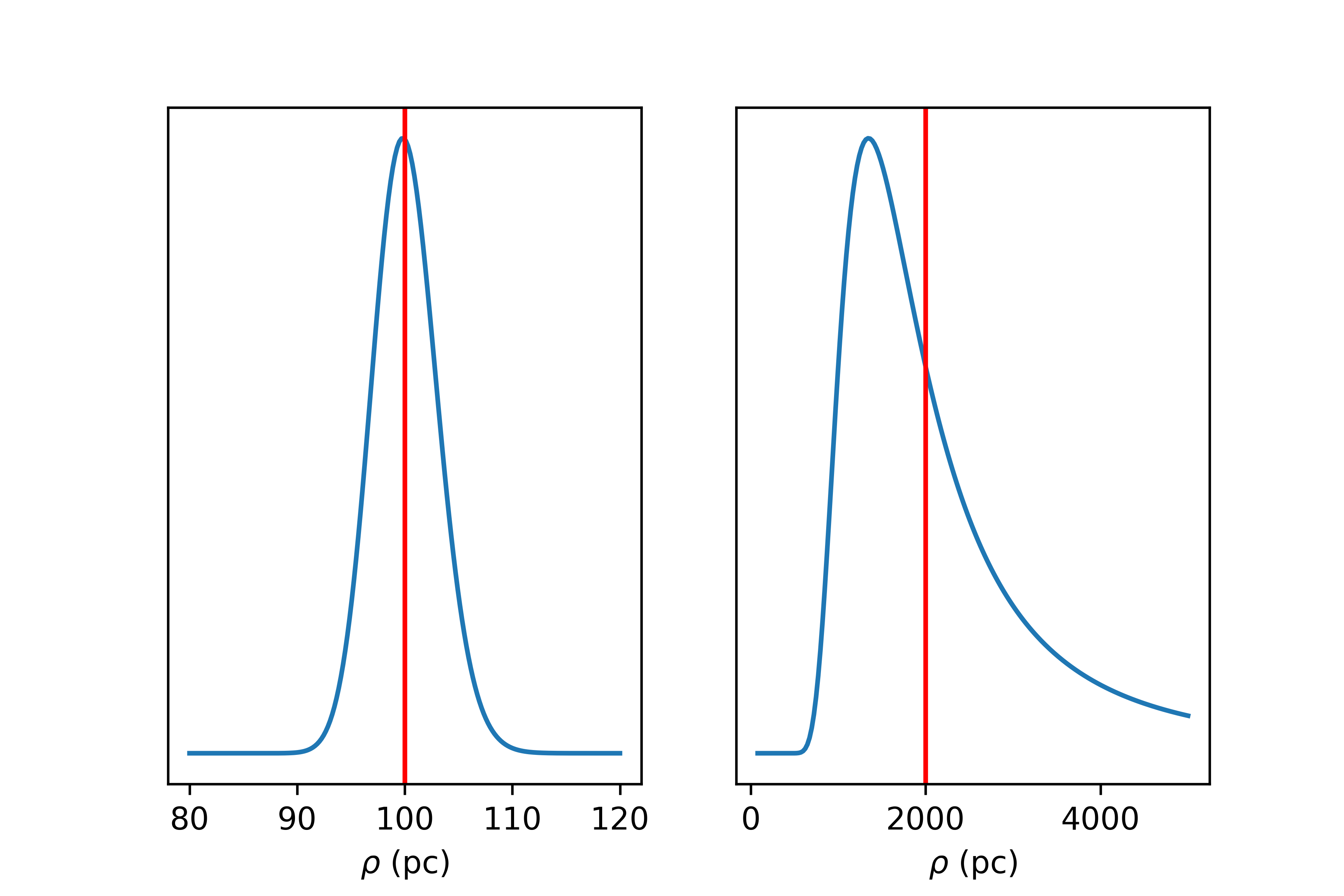}

                \caption{PDF of $\rho = 1/\parallax$ in two extreme
                  cases. The red vertical line indicates the true
                  distance $r$.  Left: Object at $r=100$ pc with an
                  uncertainty on the observed parallax of
                  $\parallaxsd=0.3$ mas. Right: Object at $r=2000$ pc                  with an uncertainty on the observed parallax of
                  $\parallaxsd=0.3$ mas.}

                \label{fig:Pdf_Rho}
\end{figure}

In \figref{fig:Pdf_Rho} we depict $p( \rho\mid\parallaxtrue )$ for
two extreme cases of very low and very high relative
uncertainty. The shape of $p( \rho\mid\parallaxtrue )$ describes what
we can expect when using $\rho$ as an estimate of the true distance
$r$. The distribution of the figure on the left corresponds to a
case with a low fractional parallax uncertainty, defined as
 $f=\parallaxsd/\parallaxtrue$. It looks unbiased and
symmetrical. Thus, using $\rho=1/\parallax$ to estimate the distance
in a case like this is relatively safe and would lead to more or less
reliable results. However, in spite of its appearance, the figure
hides an intrinsic non-Gaussianity that is made evident in the
right-hand figure. This second plot corresponds to the case of high
fractional parallax uncertainty and the distribution shows several
features: first, the mode (the most probable value) does not coincide
with the true distance value; second, the distribution is strongly
asymmetric; and finally, it presents a long tail towards large values
of $\rho$. For more extreme values of $f$
there is a noticeable negative tail to this distribution,
corresponding to the negative tail of the observed parallax
distribution.

  In view of \figref{fig:Pdf_Rho} it is tempting to apply 
  corrections to the $\rho$ estimator based on the value of
  the fractional parallax uncertainty $f$. 
        Unfortunately, in order to do so we would need to know the
  true value of the parallax and $f$. Using the
        apparent fractional uncertainty $f_{app}=\parallaxsd/\parallax$
        is not feasible since the denominator in $f$ (the
  true parallax) can be very close to zero, so its distribution
  has very extended wings and using $f_{app}$ will often result in gross
  errors.

  Furthermore, reporting a $\rho$ value should always be accompanied
  by an uncertainty estimate, usually the standard deviation of the
  estimator, but the standard deviation or the variance is defined in
  terms of an unknown quantity: $\parallaxtrue$. In addition, the long tail shown in
  the right panel of \figref{fig:Pdf_Rho} makes the estimates of the
  variance  quickly become pathological, as discussed below.

%\textcolor{blue}{AB: I've removed the paragraph and figure related to
%$\alpha$. It is not needed for the discussion and will only confuse
%matters. From here on the text should be further shortened and I
%think we don't need to emphasize the ``bias'' aspect so much. The
%naive estimator is also problematic because of the non-defined
%variance and because it can only be studied in terms of quantities
%that we have not access to (the true parallax and true relative
%error).}

%As mentioned above, Fig. \ref{fig:Pdf_Rho} shows that the most
%probable value of the distribution (its mode or maximum) does not
%coincide with the true value of the distance. In other words, if we
%were able to make many measurements of the parallax of a source, the
%most probable value of $\rho=1/\parallax$ would be a biased estimator
%of the true distance. Alternatively, we can think (as a mind
%experiment) of a set of sources all at the same distance and affected
%by the same fractional parallax uncertainty $f=0.3/0.5$ (just for illustrative purposes). The spread of
%observed parallaxes due to the observational uncertainties would
%result in estimated distances distributed according to the PDF
%depicted on the right panel of Fig. \ref{fig:Pdf_Rho}. Not only the
%mode is a biased with respect to the true value. Also the mean is
%biased in the classical sense that the expected value of the
%distribution differs from the true value.

In order to clarify the previous assertions, we recall
the classical concept of bias because it  plays a central role in
the discussion that develops in this section. In statistics, an
estimator is said to be biased if its expected value differs from the
true value. In our case, we aim to infer the true value of the
parallax $\parallaxtrue$ (or, alternatively, related quantities such as
the true distance $r$, absolute magnitude, luminosity, or 3D velocity
components), and we aim to infer it from the measured parallax.
In the {\it Gaia} case this measured parallax
will be affected by quasi-Gaussian uncertainties (see \secref{sec:uncertainities}). 
In this case the expectation
value of the observed parallax coincides with the true value:

\begin{equation}
\mathbb{E}[\parallax]=\int \parallax p(\parallax|\parallaxtrue)\cdot{\rm
  d}\parallax=\int \parallax \mathcal{N}(\parallax;\parallaxtrue,\parallaxsd)\cdot{\rm
  d}\parallax=\parallaxtrue,
\end{equation}

\noindent where $\mathcal{N}(\parallax;\parallaxtrue,\parallaxsd)$ represents
the Gaussian probability distribution centred at the true parallax and
with a standard deviation $\parallaxsd$. Hence, the observed
  parallax is an unbiased estimator of the true parallax (under the
strong hypothesis that there are no systematic biases associated with 
the survey and that the errors are normally distributed).

Now, in order to assess the bias of $\rho= 1/\parallax$ as an estimator of
the true distance we need to calculate its expected value:

\begin{equation}
\mathbb{E}[\rho]=\mathbb{E}[1/\parallax]=\int
\frac{1}{\parallax}\cdot p(\parallax|\parallaxtrue)\cdot{\rm
  d}\parallax=
\int
\frac{1}{\parallax}\cdot\mathcal{N}(\parallaxtrue,\parallaxsd)\cdot{\rm
  d}\parallax
  \end{equation}

\begin{figure*}[htb]%textwidth
        \centering
        \includegraphics[width=0.33\textwidth]{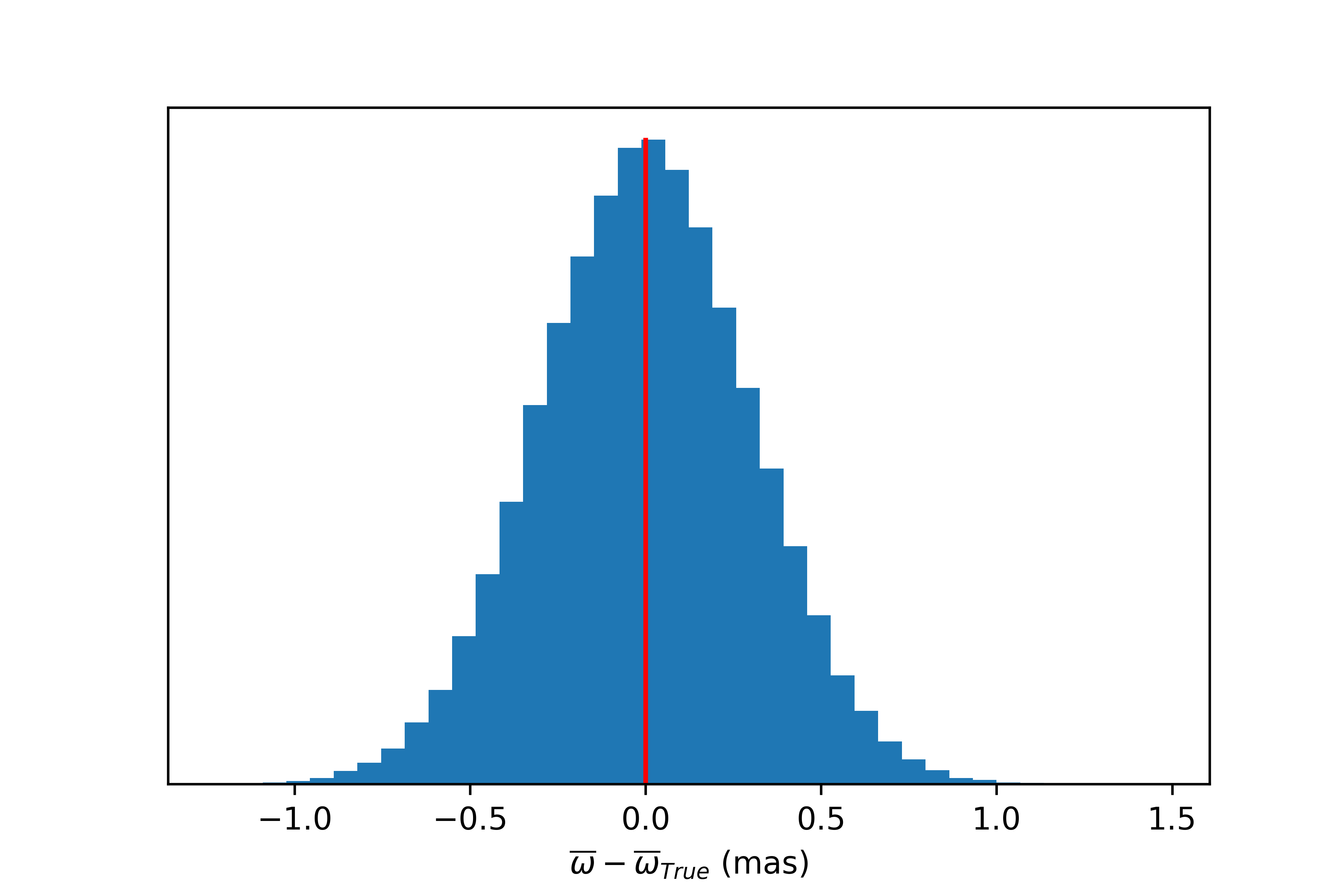}
        \includegraphics[width=0.33\textwidth]{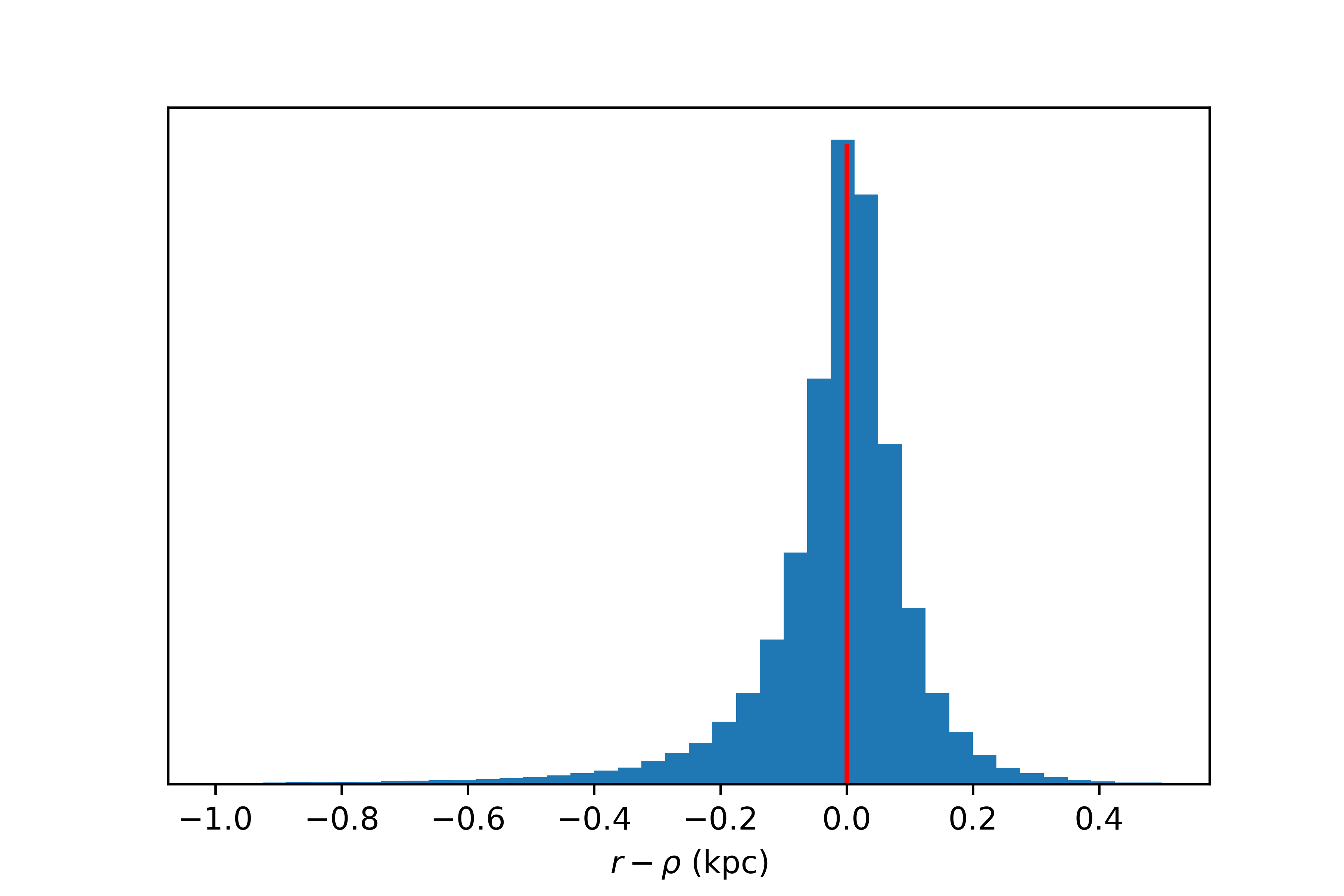}
        \includegraphics[width=0.33\textwidth]{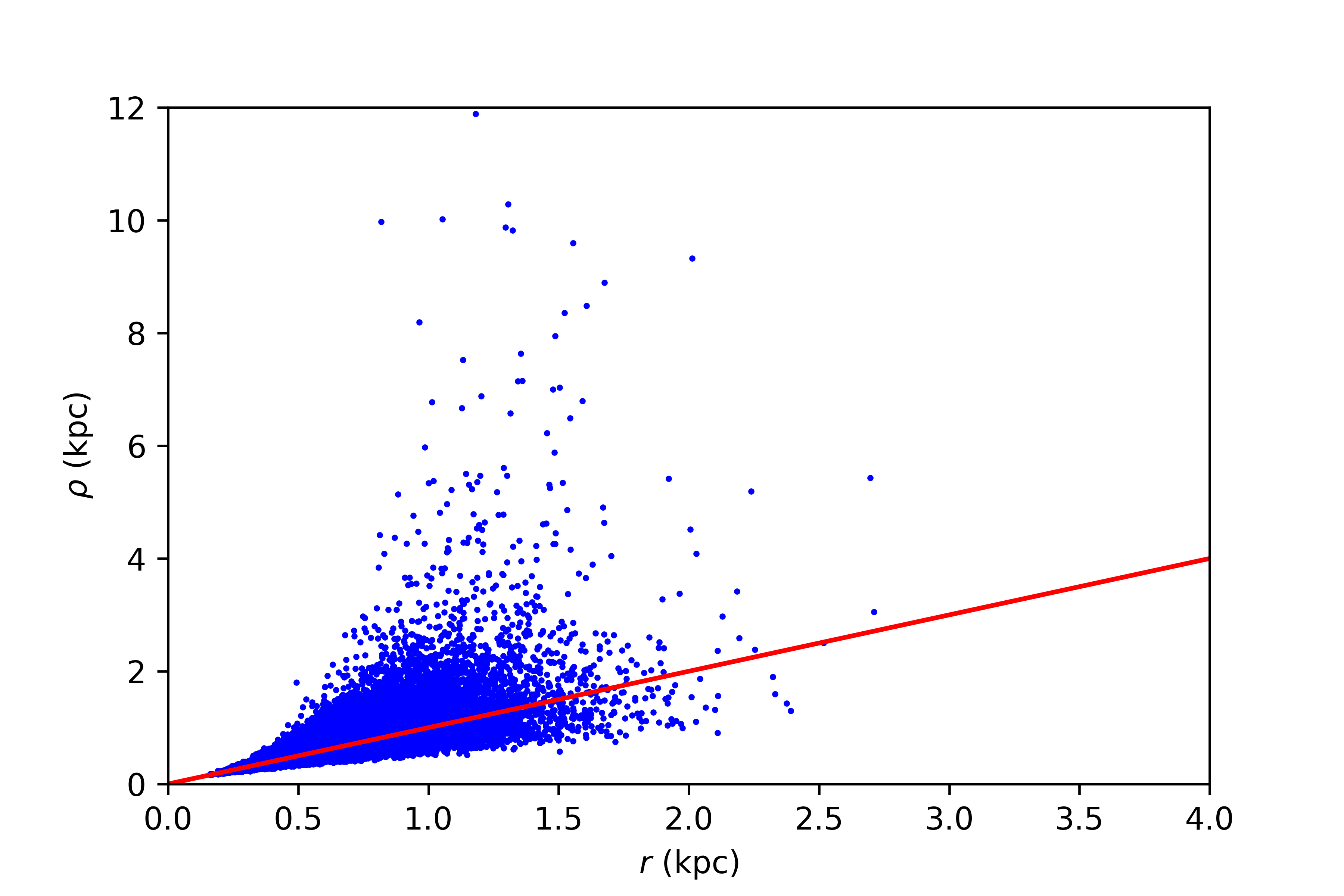}
        \caption{Behaviour of PDF of $\rho = 1/\parallax$ as estimator
          of the true distance.  Left: Histogram of differences
          between true parallaxes and observed parallaxes.  Centre:
          Histogram of differences between true distances and their
          estimation using $\rho$.  Right: Comparison of the true
          distances and their estimations using $\rho$.  The observed
          parallaxes $\parallax$ have been simulated using an
          uncertainty of $\parallaxsd=0.3$ mas.}
        \label{fig:r_vs_rho}
\end{figure*}

\noindent This bias was approximated by \cite{SmithEichhorn} (see
\secref{sec:Smith-Eichhorn}) as a function of the fractional parallax
uncertainty $f$ using a series expansion of the term in the integral
and several approximations for the asymptotic regimes of small and
large values of $f$, and it indeed shows  that the distance estimator
$1/\parallax$ is unbiased for vanishingly small values of $f$, but it
rapidly becomes significantly biased for values of $f$ beyond 0.1.
But not only is $1/\parallax$ a biased estimator of the true
distance,  it is also a high-variance estimator. The reason for this
variance explosion is related to the long tail towards large distances
illustrated in the right panel of \figsref{fig:Pdf_Rho} and
\ref{fig:r_vs_rho}. Relatively large fractional uncertainties
inevitably imply noise excursions in the parallax that result in
vanishingly small observed parallaxes and disproportionate distances
(and hence an inflation of the variance). 

The effects discussed above can be illustrated with the use of
simulated data. \figrefalt{fig:r_vs_rho} shows
the results of a simulation of objects located between 0.5 and 2kpc
where starting from the true distances we have simulated observed
parallaxes with a Gaussian uncertainty of $\parallaxsd=0.3$ mas and
then calculated for each object $\rho = 1/\parallax$.

The figure on the left shows that (by construction) the errors in the
observed parallaxes are well behaved and perfectly symmetrical (Gaussian),
while in the centre figure the errors in the estimation of distances
using $\rho$ show a strong asymmetry. The characteristics of these
residuals depend on the distribution of true distances and
uncertainties. This is more evident in the figure on the right, where
the true distance $r$ is plotted against $\rho$; there is a very prominent
tail of overestimated distances and the distribution is asymmetrical
around the one-to-one line:  the more distant  the
objects, the more marked the asymmetry. These features are very prominent because we have simulated
objects so that the relative errors in parallax are large, but they
are present (albeit at a smaller scale) even when the relative errors
are small.

\begin{figure*}[htb]
        \centering
        \includegraphics[width=1.00\textwidth]{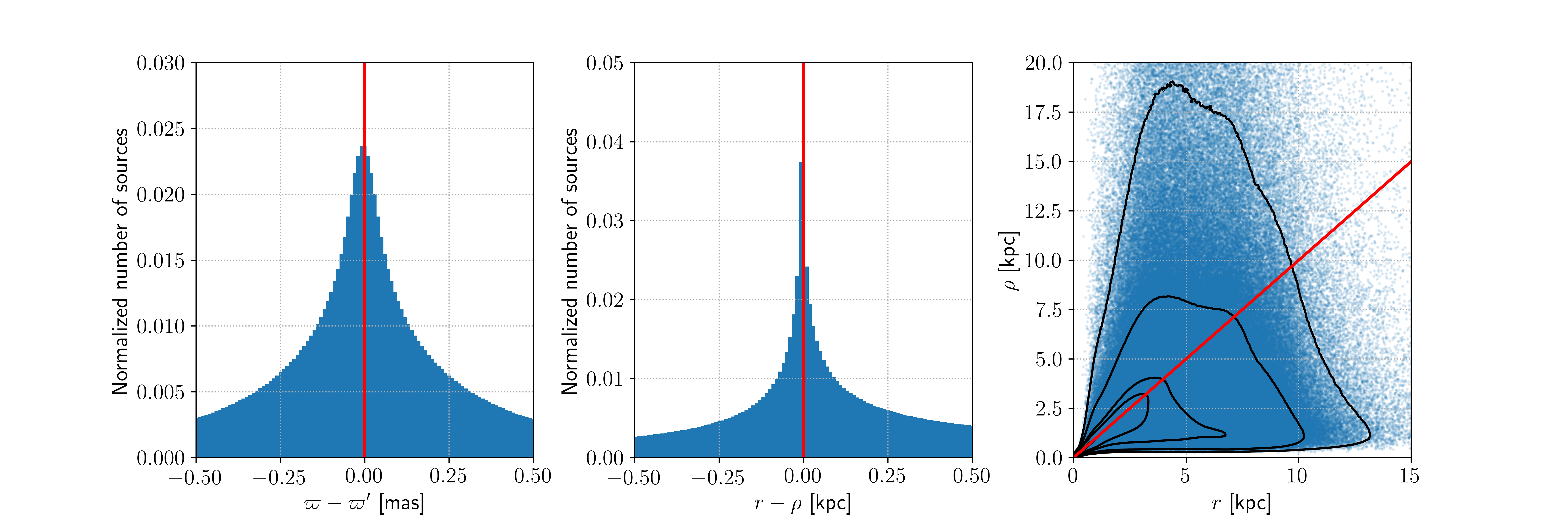}
        \caption{Behaviour of PDF of $\rho = 1/\parallax$ as estimator
          of the true distance for a simulation of the full \gdrtwo
          data set.  Left: Histogram of differences between true
          parallaxes and observed parallaxes.  Centre: Histogram of
          differences between true distances and their estimation
          using $\rho$.  Right: Comparison of the true distances and
          their estimations using $\rho$.  The observed parallaxes
          $\parallax$ have been simulated using a realistic \gdrtwo error          model described in the Appendix. The contour lines 
                                        correspond to the distribution percentiles of 35\%, 55\%, 
                                        90\%, and 98\%.}
        \label{fig:r_vs_rho-DR2}
\end{figure*}

The plots in \figref{fig:r_vs_rho} correspond to a simple
simulation with a mild uncertainty $\parallaxsd=0.3$ mas. 
\figrefalt{fig:r_vs_rho-DR2} shows the same plots for a realistic simulation
of the \gdrtwo data set.  The simulation is described in  Appendix
\ref{sec:app_samples}; in this case the errors in parallax follow a
realistic model of the \gdrtwo errors, depicted in \figref{fig:errors_g}.

%\textcolor{blue}{We have to mention/analyse the global Bias that
%would result to calculate 1/pi distances for the whole DRe. Even if
%small it is there, and becomes relevant for large relative errors.}

As a summary, we have seen in previous paragraphs that the naive
approach of inverting the observed parallax has significant
drawbacks: we are forced to dispose of valuable data (non-positive parallaxes),
and as an estimator $\rho=1/\parallax$ is biased and has a very high variance.

%%%%%%%%%%%%%%%%%%%%%%%%%%%%%%%%%%%%%%%%%%%%%%%%%%%%
\subsection{Sample truncation}
  \label{sec:truncation}

In addition to the potential sources of trouble described 
in the previous sections, the traditional use of samples of 
parallaxes includes a practice that tends to aggravate 
these effects: truncation of the used samples. 

As discussed in \secref{sec:measurement}, negative parallaxes are a natural
result of the \gaia measurement process (and of astrometry
in general). Since inverting negative parallaxes leads 
to physically meaningless negative distances we are  tempted
to just get rid of these values and form a `clean' sample.
This results in a biased sample, however. 

On the one hand, removing the negative parallaxes biases the
distribution of this parameter. Consider for instance the
case illustrated in \figref{fig:QSO_Normalized_Parallaxes}
for the quasars from the AllWISE catalogue. These objects have
a near zero true parallax, and the distribution of its observed values
shown in the figure corresponds to this, with a mean of $-10$~$\mu$as,
close to zero. However, if we remove the
negative parallaxes from this sample, deeming them `unphysical',
the mean of the observed values would be significantly positive,  
 about $0.8$ mas. This is completely unrealistic for quasars; 
in removing the negative parallaxes we have significantly biased 
the observed parallax set for these objects. With samples of
other types of objects with non-zero parallaxes the effect can be
smaller, but it will be present.

\begin{figure}[htb]
        \centering
                \includegraphics[width=\columnwidth]{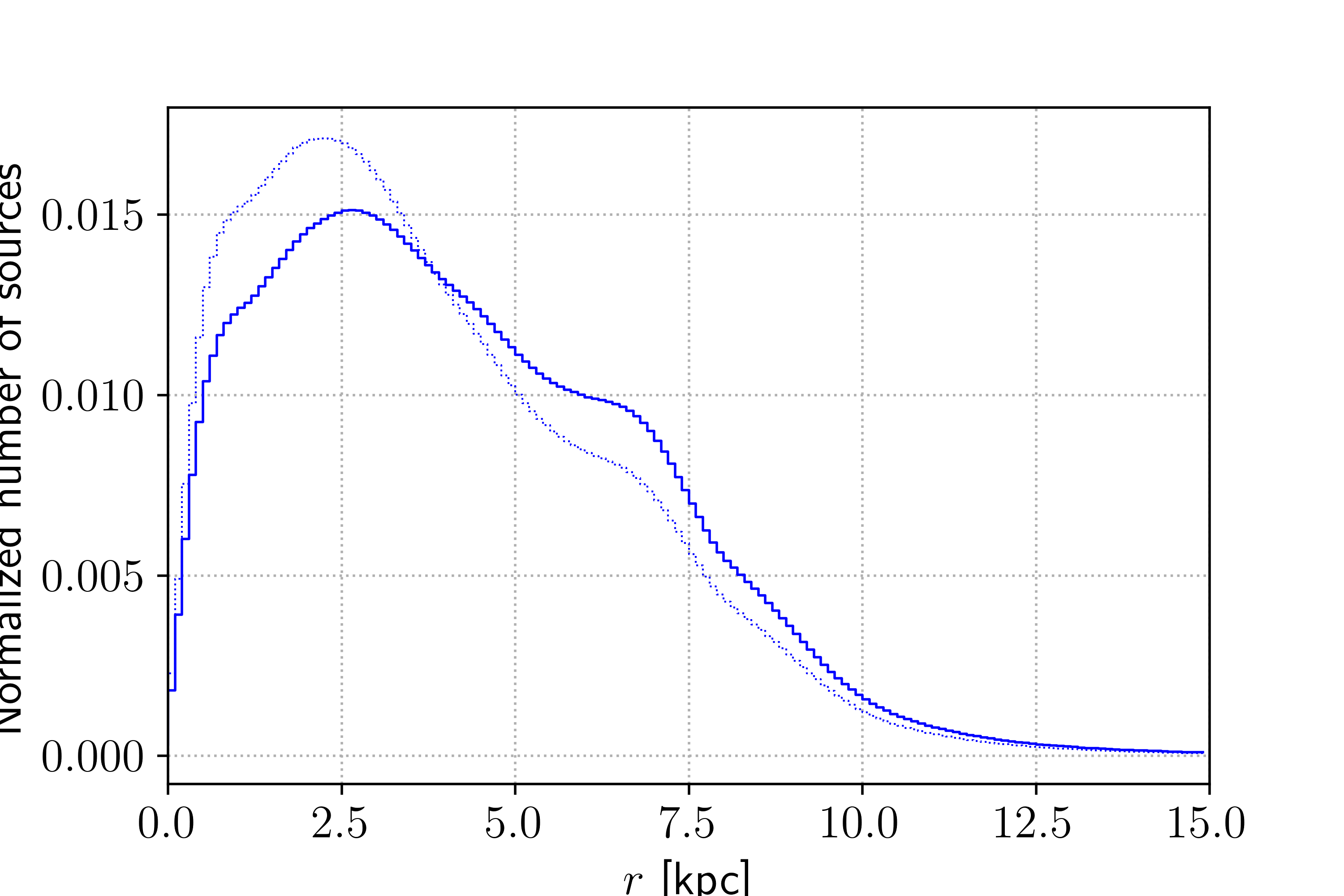}
          \caption{Effect of removing the negative and zero parallaxes from a simulation of \gdrtwo.
                                                 Distribution of true distances:
                                                 histogram of distances for the complete sample (thick line),                                                  histogram of distances for the sample truncated by removing $\parallax \leq 0$ (thin line).
                                        }
          \label{fig:Hist_R_bias_neg_pi}
  \end{figure}

On the other hand, when by removing negative parallaxes
the contents of the sample are no longer representative of the base population
from which it has been extracted since stars with large parallaxes
are over-represented and stars with small parallaxes are under-represented.
This can be clearly illustrated using a simulation. We have generated a
sample of simulated stars mimicking the contents of the full \gdrtwo (see Appendix \ref{sec:app_samples}) and truncated it by removing the negative
parallaxes. In \figref{fig:Hist_R_bias_neg_pi} we can compare the
distribution of the true distances of the original (non-truncated) sample and the resulting (truncated) sample; it is clear that after the
removal of negative parallaxes we have favoured the stars at short
distances (large parallaxes) with respect to the stars at large 
distances (small parallaxes). The spatial distribution of the sample
has thus been altered, and may therefore bias any analysis based
on it.

\begin{figure}[htb]
        \centering
                \includegraphics[width=1.00\columnwidth]{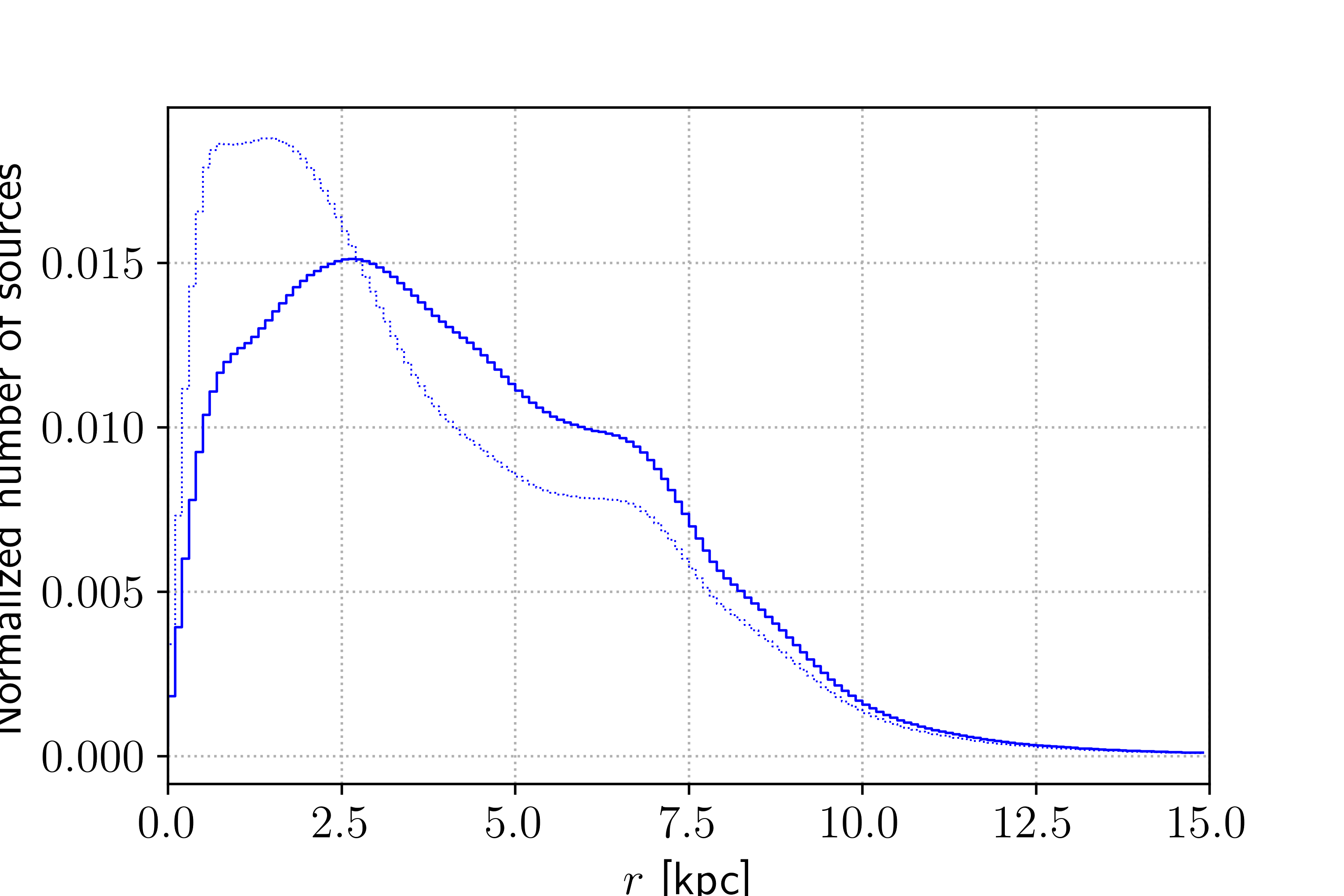}
          \caption{Effect of removing the positive parallaxes with a relative error above 50\% as well as negative parallaxes.
                                                 Thick line: histogram of distances for the complete sample.
                                                 Thin line: histogram of distances for the sample truncated by removing objects with ($|\frac{\parallaxsd}{\parallax}| > 0.5$).
                                        }
          \label{fig:Hist_R_bias_50percent_pi}
  \end{figure}

A stronger version of truncation that has traditionally been applied is to remove not only negative parallaxes, but also all the parallaxes with
a relative error above a given threshold $k$, selecting $\frac{\parallaxsd}{\parallax}<k$. 
This selection tends to favour the removal of stars with small parallaxes. The effect is similar to the
previous case, but more accentuated as  can be seen in \figref{fig:Hist_R_bias_50percent_pi}.
Again, stars at short distances are favoured in the sample with respect
to distant stars.

\begin{figure}[htb]
        \centering
                \includegraphics[width=\columnwidth]{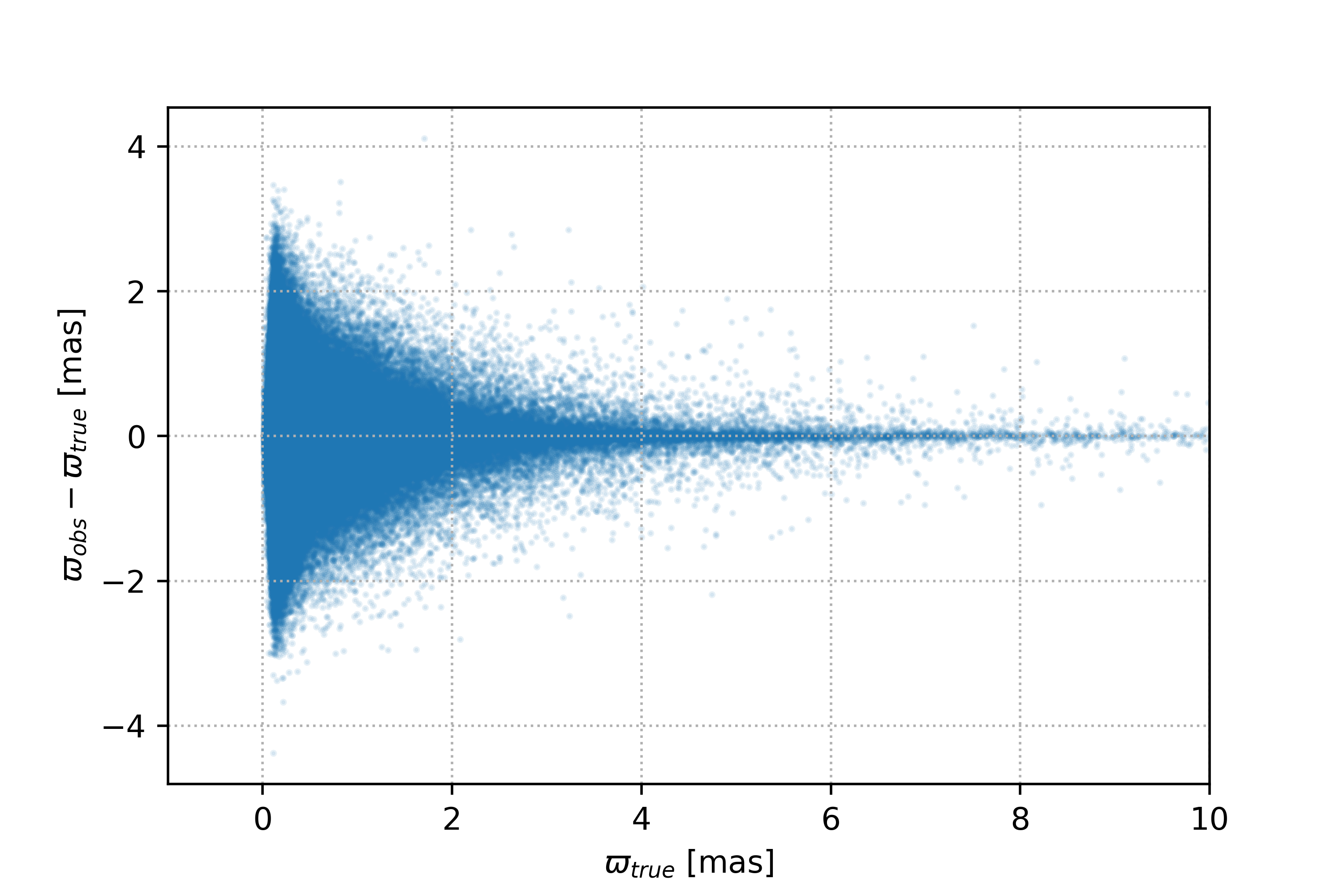}
                
                \includegraphics[width=\columnwidth]{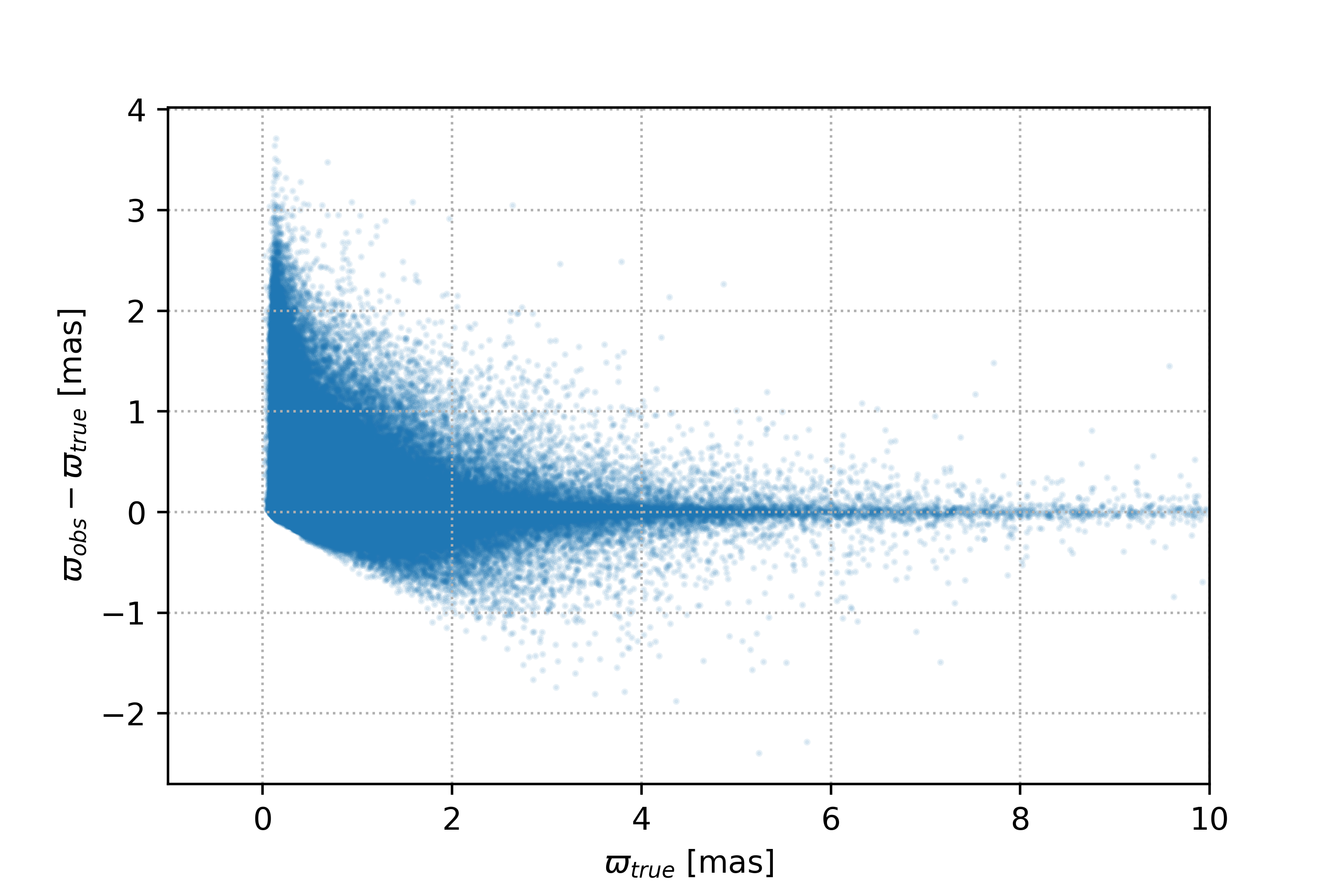}
                
                \includegraphics[width=\columnwidth]{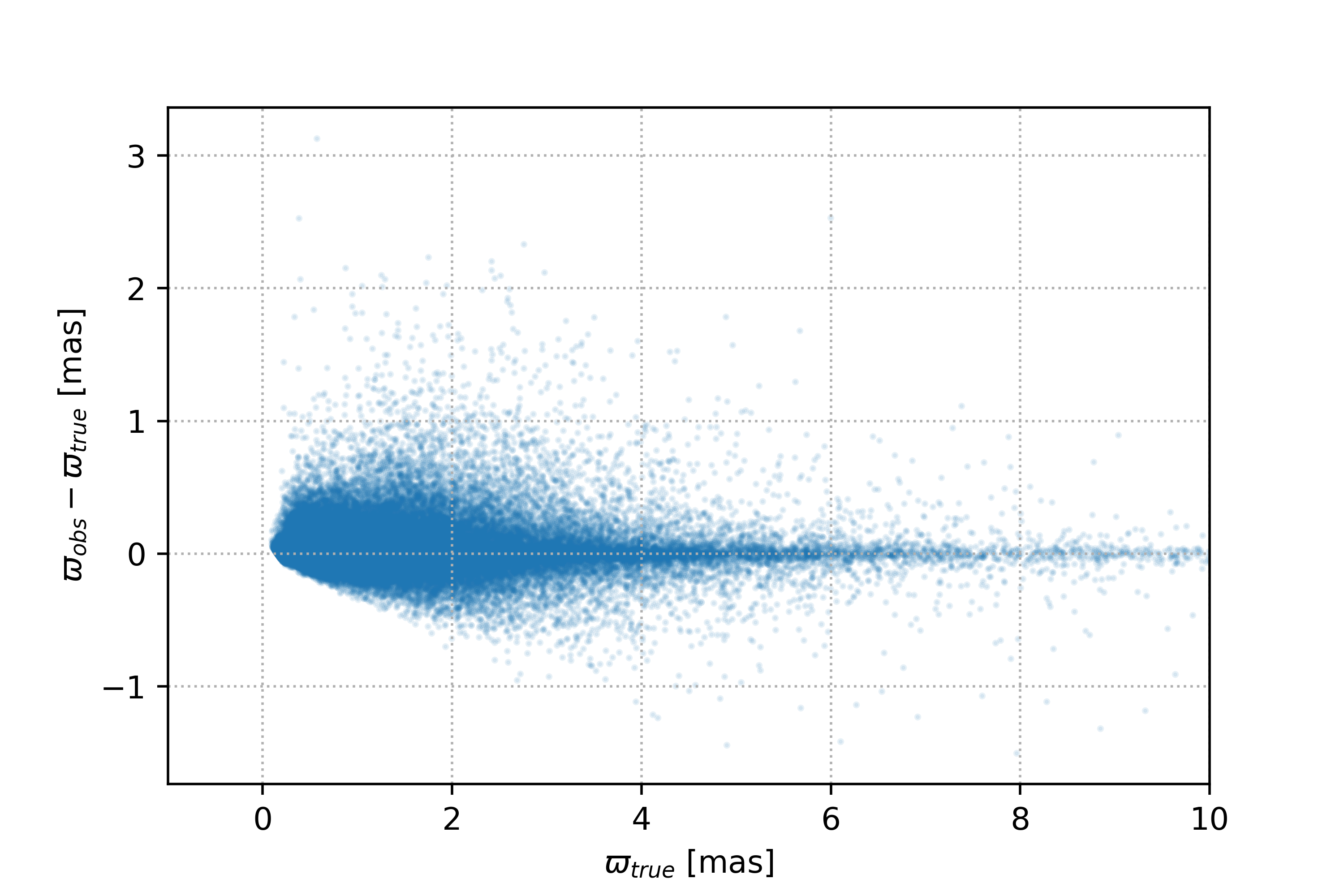}
                \caption{Differences between true and observed parallax: effect of removing observed negative parallaxes and those above a given relative error. We start with a representative subsample of 1 million stars (top figure) and truncate it according to the apparent relative parallax precision. 
                                                 {\bf Top:} Complete sample. Mean difference $\parallax - \parallaxtrue$ is $1.55 \times 10^{-5}$ mas.
                                                 {\bf Middle:} Retaining only objects with positive parallaxes and $|\frac{\parallaxsd}{\parallax}| < 0.5$. The mean difference $\parallax - \parallaxtrue$ is $0.164$ mas \hspace{0.25cm}.
                                                 {\bf Bottom:} Retaining only objects with positive parallaxes and $|\frac{\parallaxsd}{\parallax}| < 0.2$. The mean difference $\parallax - \parallaxtrue$ is $0.026$ mas.
                                        }
          \label{fig:Diff_true_obs_pi}
  \end{figure}

Even worse, as in the previous case  the  truncation  not only makes the
distribution of true distances unrepresentative, but it  also biases 
the distribution of observed parallaxes: stars with 
positive errors (making the observed parallax larger than the true one) tend to 
be less removed than stars with negative errors (making the observed
parallax smaller than the true one). By favouring positive
errors with respect to negative errors, we are also biasing the
overall distribution of parallaxes. \figrefalt{fig:Diff_true_obs_pi}
depicts this effect. The plots show the
difference $\parallax - \parallaxtrue$ as a function of $\parallaxtrue$.
We can see in the middle and bottom figures how the removal of
objects is non-symmetrical around the zero line, so that the overall
distribution of $\parallax - \parallaxtrue$ becomes biased. From 
an almost zero bias for the full sample (as expected from \gaia in
absence of systematics) we go to significant biases once we 
introduce the truncation,  and the bias is dependent of the cut 
value we introduce.

Furthermore, in {\it Gaia} the parallax uncertainties vary with the
object magnitude, being larger for faint stars (see \figref{fig:errors_g}).
Therefore, a threshold on the relative error will favour bright stars
over the faint ones, adding to the above described biases.

\begin{figure}[htb]
        \centering
                \includegraphics[width=1.00\columnwidth]{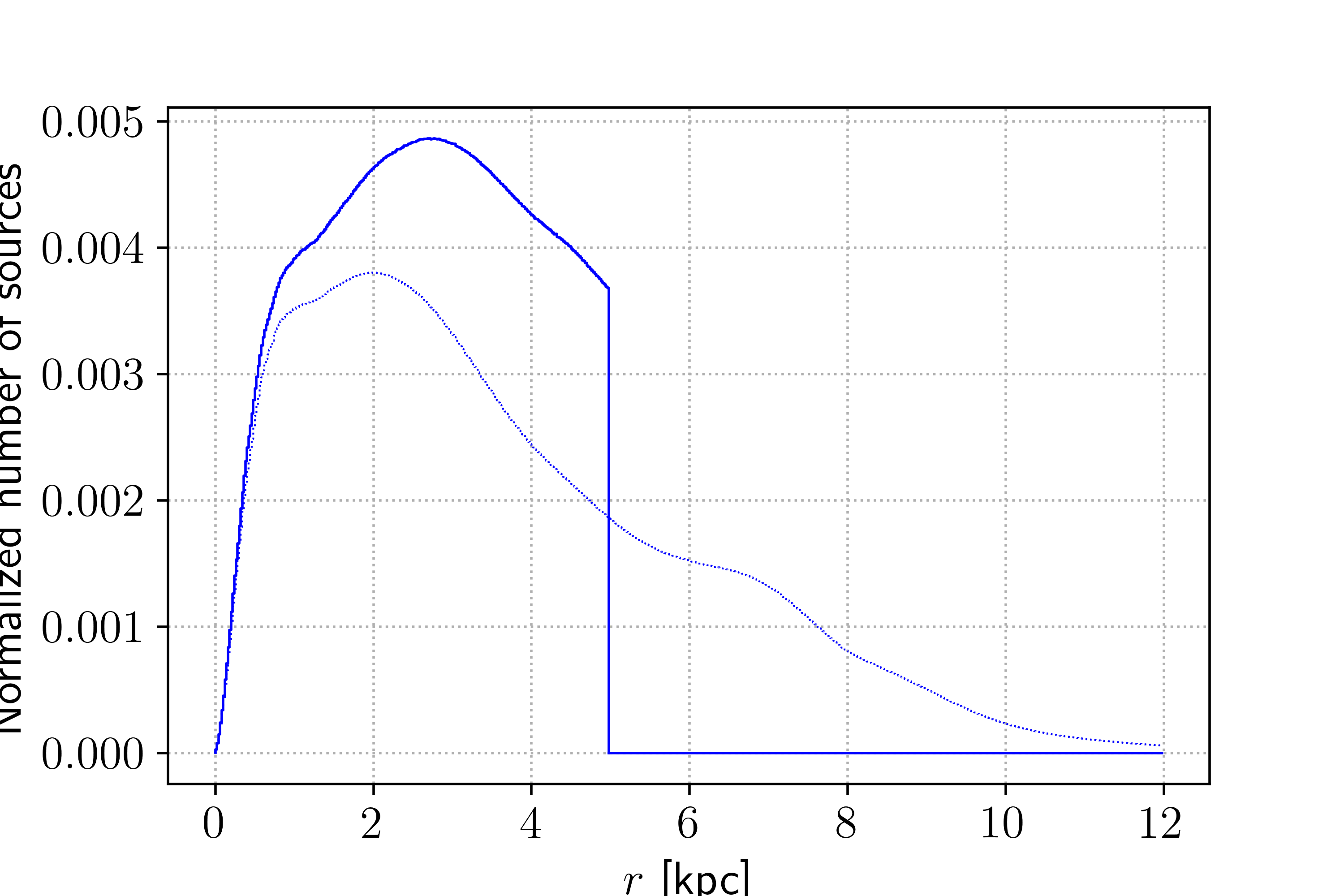}
          \caption{Effect of introducing a limit on the observed parallax.   
                                                 Thick line: histogram of distances up to 5 kpc for the complete sample.
                                                 Thin line: histogram of distances for the sample truncated by removing objects with $\parallax < 0.2$ mas 
                                                   (distance estimated as $1/\parallax$ up to 5 kpc).
                                        }
          \label{fig:Hist_R_bias_limit_5kpc}
  \end{figure}

Another type of truncation that has been traditionally applied 
is to introduce limits in the observed parallax. The
effects of such a limit are closely related to the Lutz--Kelker
bias discussed in \secref{sec:Lutz_Kelker}. Suffice it here
to illustrate the effect with a specific example on a \gdrtwo-like 
sample. If we take the full sample and remove stars with 
$\parallax < 0.2$ mas we could imagine that we are roughly 
removing objects further away than 5 kpc. However, the net result
is depicted in \figref{fig:Hist_R_bias_limit_5kpc} where
we can see that instead of the distribution of true distances of the
complete sample up to 5 kpc (solid line) we get a distribution
with a lack of closer stars and a long tail of stars with
greater distances. A larger limit in parallax (shorter limiting distance) will produce
a less prominent effect since the relative errors in the parallax
will be smaller, but the bias will be nonetheless present.

In conclusion, our advice to readers is to avoid introducing
truncations when using {\it Gaia} data since, as illustrated above,
they can strongly affect the properties of the sample and therefore
affect the data analysis. If  truncation is unavoidable
it should be included in the Bayesian modelling of the overall
problem (see \secref{sec:summary_recommendations}).

%%%%%%%%%%%%%%%%%%%%%%%%%%%%%%%%%%%%%%%%%%%%%%%%%%%%
\subsection{Corrections and transformations}

%We have seen in previous paragraphs that the naive approach of
%inverting the observed parallax brings undesired effects (). First of
%all, we are forced to dispose of valuable data (negative parallaxes)
%which carry perfectly valid information. As a consequence of this
%sample truncation we are introducing biases because not all sources are
%equally susceptible to yield negative parallaxes. Even for sources
%with positive parallaxes, the non-linear transformation from observed
%parallax to distance translates into a systematic difference between
%the mode or the mean of $p(\rho|\parallax)$) and the true value (a
%bias of the mode and the mean if used as distance estimators). Finally, this
%non-linear relation also results in an explosion of the variance due
%to the arbitrarily small values of the observed parallax that can be
%measured when the fractional parallax uncertainty is large.

%As we will see, this is not the only source of bias in parallax data
%sets. There have been some attempts to circumvent these and other
%problems associated with the inversion of the observed parallax by
%introducing {\sl ad hoc} corrections.

In this section, we review proposals in the literature for the
use of parallaxes to estimate distances. In general they take the
form of `remedies' to correct one or another problem on this use.
Here we explain why these remedies cannot be recommended.

\subsubsection{The Smith--Eichhorn correction}
\label{sec:Smith-Eichhorn}

\cite{SmithEichhorn} attempt to compensate for the bias introduced by the naive
inversion of the observed parallax (and the associated variance
problem) in two different ways. The first  involves transforming
the measured parallaxes into a pseudo-parallax $\varpi^*$ according to

\begin{equation}
 \varpi^* \equiv
 \beta\cdot\parallaxsd\left[\frac{1}{\exp(\phi)+\exp(\frac{-1.6\parallax}{\parallaxsd})}+\phi\right]
\label{eq:smith1}
,\end{equation}

\noindent where
$\phi\equiv\ln(1+\exp(\frac{2\varpi}{\sigma_{\varpi}}))/2$ and $\beta$
is an adjustable constant. The qualitative effect of the
transformation is to map negative parallaxes into the positive
semi-axis $\mathbb{R}^+$ and to increase the value of small parallaxes
until it asymptotically converges to the measured value for large
$\varpi$. For $\varpi=0$, $\phi =\ln(2)/2$ and the pseudo-parallax
value
$\varpi^*=\beta\cdot\sigma_{\varpi}\left(\frac{1}{1+\exp(\ln(2)/2)}+\frac{\ln(2)}{2}\right)$
has the undesirable property of depending on the choice of $\beta$ and on
the parallax uncertainty. Thus, even in the case of a small parallax
measurement $\varpi\rightarrow 0$ with a relatively small fractional
parallax uncertainty (e.g. $f=0.1$), we 
substitute a perfectly useful and valuable measurement by an arbitrary
value of $\varpi^*$.

%The transformation includes a free parameter ($\beta$) that needs to
%be fixed before-hand and several {\sl ad hoc} constants (-1.6 and 2)
%that have been fixed by trial-and-error. Unfortunately, the
%trial-and-error procedure is not specified by the authors and hence it
%is not known under what conditions the bias reduction applies and to
%what extent it is of general applicability.

%According to \cite{SmithEichhorn}, the new pseudo-parallax follows a
%distribution with the desired statistical property that the expected
%bias is significantly reduced with respect to the direct inversion of
%the measured parallaxes. This proposal can be applied to individual
%measurements and Eq. \ref{eq:naiveInversion} used to yield individual
%distances. 

%The bias reduction is a direct consequence of the remapping of the
%observed parallaxes that are negative or close to zero. These
%parallaxes get artificially mapped to larger values of the
%pseudo-parallax thus avoiding the implausible large (or even negative)
%distances.

The Smith--Eichhorn transformation is an arbitrary (and rather
convolved) choice amongst many such transformations that can reduce
the bias for certain particular situations. Both the analytical
expression and the choice of constants and $\beta$ are the result of
an unspecified trial-and-error procedure, the applicability of which
is unclear. Furthermore, as stated by the authors, they introduced a
new bias because the transformed parallax is always larger than the
measured parallax. This new bias has no physical interpretation
because it is the result of an {ad hoc} choice for the analytical
expression in \equref{eq:smith1}. It is designed to reduce the bias,
but it does so by substituting perfectly reasonable direct
measurements (negative and small parallaxes) that we can interpret and
use for inference, by constructed values arising from the choice.

\subsubsection{The Lutz--Kelker correction}
\label{sec:Lutz_Kelker}

\begin{figure*}[htb]
        \centering
        \includegraphics[width=1.00\textwidth]{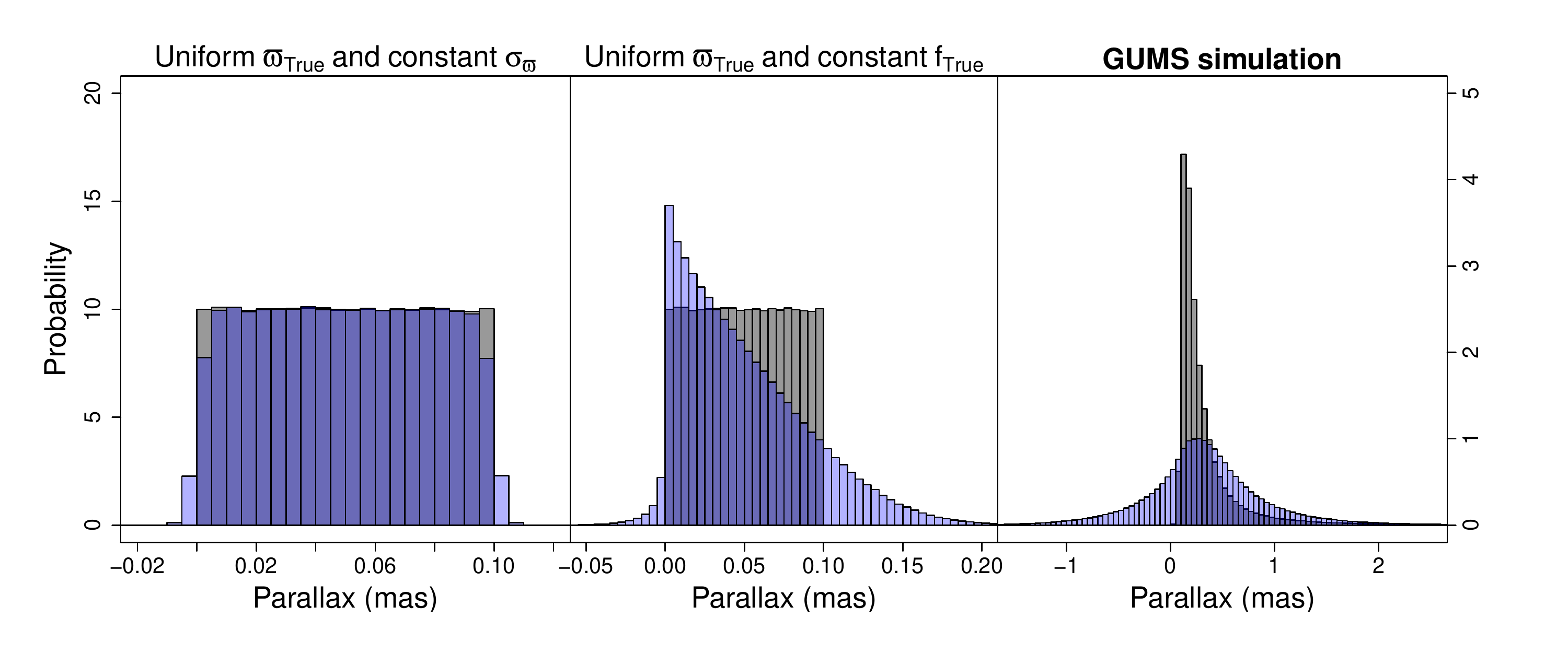}
        \caption{Histograms of true (grey) and observed (blue)
          parallaxes for three simulations: uniform distribution of
          parallaxes and constant $\parallaxsd=0.3$\ mas (left);
          uniform distribution of parallaxes and constant $f=0.5$ (centre); and the GUMS simulation described in
          Appendix \ref{sec:app_samples} (right).}
        \label{fig:LK-TE1}
\end{figure*}

\cite{LutzKelker73} realised that the spatial distribution of sources
around the observer together with the unavoidable observational errors and a
truncation of the sample on the value of the observed parallax result in
systematic biases in the average parallax of certain stellar samples. The bias described by \cite{LutzKelker73} is a manifestation of the truncation
biases described above and can be understood if we look at a few very simple
examples. First, let us imagine a density of sources around the
observer such that the distribution of true parallaxes
$p(\parallaxtrue)$ is constant in a given interval and zero
outside. Let us also imagine that the observation uncertainty
$\parallaxsd$ is constant and equal to 0.3 mas. The left panel of
\figref{fig:LK-TE1} shows the distribution of true and observed
parallaxes for a simulation of such a situation and $10^6$ sources. We
see that the observed parallaxes are also approximately uniform and
the departures from uniformity appear near the edges. For a given bin
of intermediate parallax we have as many sources contaminating from
neighbouring bins as we have  sources lost to other bins due to the
observational uncertainties, and the result is a negligible net flux.

In the middle panel we show the same histograms except that instead of
a constant parallax uncertainty, we use a constant value of the
fractional parallax uncertainty $f$. This is still an unrealistic
situation because the distribution of true parallaxes is not uniform
and also because a constant $f$ implies that larger parallaxes are
characterised by larger uncertainties. However,  it helps us to illustrate
that even in the case of uniform true parallaxes we may have a
non-zero net flux of sources between different parallax bins depending
on the distribution of parallax uncertainties. In this case, noise
shifts the value of large parallaxes more than that of smaller parallaxes (because of the constant value of $f$), so larger parallaxes get more
scattered, thus suppressing the distribution more at large
parallaxes.

Finally, the right panel shows the same plot for a realistic
distribution of distances from a  \gaia Universe Model
Snapshot (GUMS) sample (see Appendix \ref{sec:app_samples} for a full description). We see that the effect of a realistic non-uniform distribution of
parallaxes and parallax uncertainties results in a net flux in the
opposite direction (smaller true parallaxes become more suppressed and
larger parallax bins are enhanced; in both cases, the bins of negative
parallaxes become populated). This is the root of the Lutz--Kelker
bias. It is important to distinguish between the Lutz--Kelker bias and
the Lutz--Kelker correction. The Lutz--Kelker bias is the negative
difference for any realistic sample between the average true parallax
and the average measured parallax (i.e. the average true parallax
is smaller than the average measured parallax). This bias has been known at
least since the work of \cite{1953stas.book.....T}, although it was
already discussed in a context different from parallaxes as early as
\cite{1913MNRAS..73..359E}. The Lutz--Kelker correction presented in
\cite{LutzKelker73} and discussed below is an attempt to remedy this
bias based on a series of assumptions.

%The number of sources in the bins with positive parallaxes close to
%zero is much smaller in the histogram of observed parallaxes (blue)
%than in the histogram of true parallaxes (grey). The sources in these
%bins are observed at negative parallaxes or contaminate the bins of
%larger parallaxes. The situation is the opposite in the bins of
%parallaxes larger than $\approx$ 0.5 mas where the number of sources
%in the histogram of observed parallaxes is significantly larger than
%in the histogram of true parallaxes because of the net influx from
%bins with smaller true parallaxes. Any bin of observed parallaxes is
%preferentially contaminated by sources with smaller true parallaxes.

In the case of Gaussian uncertainties such as those described in
\equref{eq:pdf_truepi} it is evident that the probability of
measuring a value of the parallax greater than the true parallax
$p(\parallax > \parallaxtrue|\parallaxtrue)=0.5$. The same value holds
for the probability that $\parallax < \parallaxtrue$ because the
Gaussian distribution is symmetrical with respect to the true value of
the parallax. This is also true for the joint probability of
$\parallax$ and $\parallaxtrue$,

\begin{equation}
  p(\parallax > \parallaxtrue)  = \iint_\mathcal{S} p(\parallax,\parallaxtrue)\cdot{\rm d}\parallax\cdot{\rm d}\parallaxtrue
  =0.5
  ,\end{equation}

\noindent where $\mathcal{S}$ is the region of the $(\parallax,\parallaxtrue)$
plane where $\parallax > \parallaxtrue$.

However, the probability distribution of the true parallax given the
observed parallax $p(\parallaxtrue|\parallax)$ does not fulfil this
seemingly desirable property of probability mass equipartition at the
$\parallax=\parallaxtrue$ point. We can write the latter probability
as

\begin{equation}
  p(\parallaxtrue|\parallax)=\frac{p(\parallax,\parallaxtrue)}{p(\parallax)}=
  \frac{p(\parallax|\parallaxtrue)\cdot p(\parallaxtrue)}{p(\parallax)}
\label{eq:LKposterior}
\end{equation}

\noindent using the product rule of probability. In the context of
inferring the true parallax from the observed one,
\equref{eq:LKposterior} is the well-known Bayes' theorem, where the
left-hand side is the posterior probability,
$p(\parallax\mid\parallaxtrue)$ is the likelihood, $p(\parallaxtrue)$
is the prior, and $p(\parallax)$ is the evidence. For most realistic
prior distributions $p(\parallaxtrue)$, neither the median nor the
mode or the mean of the posterior in \equref{eq:LKposterior} is at
$\parallax=\parallaxtrue$. Let us take for example a uniform volume
density of sources out to a certain distance limit. In such a
distribution the number of sources in a spherical shell of
infinitesimal width at a distance $r$ scales as $r^2$, as does the
probability distribution of the distances. Since

\begin{equation}
p(r)\cdot{\rm d}r=p(\parallaxtrue)\cdot {\rm d}\parallaxtrue,
\end{equation}

\noindent the probability distribution for the true parallax in such a truncated
constant volume density scenario is proportional to

\begin{equation}
p(\parallaxtrue) \propto \parallaxtrue^{-4}
\end{equation}

\noindent out to the truncation radius. Hence, for Gaussian distributed
uncertainties we can write $p(\parallaxtrue|\parallax)$ as

\begin{equation}
p(\parallaxtrue|\parallax) \propto \frac{1}{\sigma_{\varpi}}\cdot
\exp(\frac{-(\parallax-\parallaxtrue)^2}{2\sigma_\varpi})\cdot\parallaxtrue^{-4}.
\label{eq:LKposterior2}
\end{equation}

\begin{figure}[htb]
        \centering
        \includegraphics[width=1.00\columnwidth]{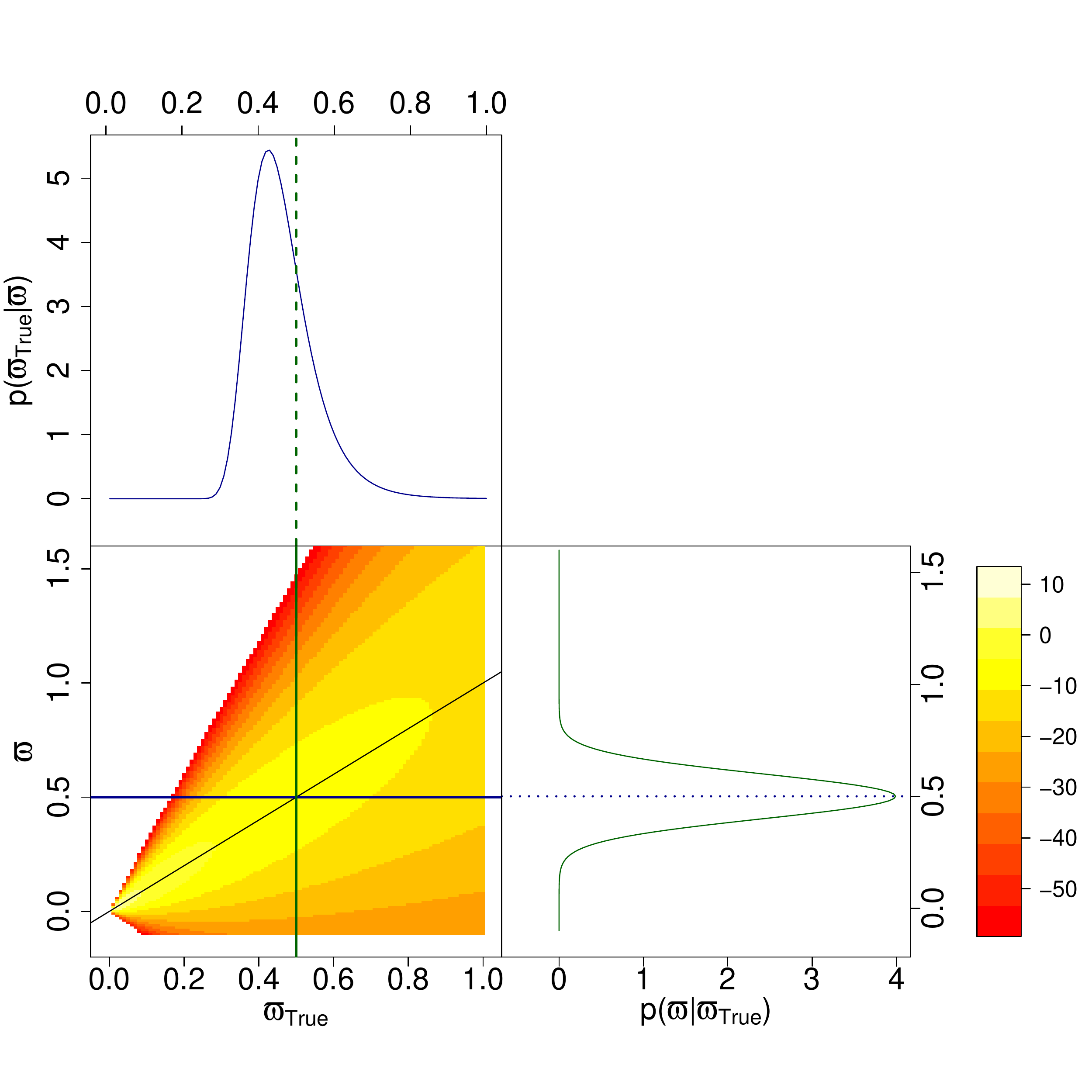}
        \caption{(Lower left) Joint probability distribution (in
          logarithmic scale to improve visibility) for the random
          variables $\parallax$ and $\parallaxtrue$ in the scenario of
          a truncated uniform volume density. The colour code is shown
          to the right of the lower right panel, and white marks the
          region where the probability is zero. (Lower right)
          Conditional probability distribution of the observed
          parallax for $\parallaxtrue$=0.5. (Upper left)
          Conditional probability distribution for $\parallaxtrue$
          given an observed parallax $\parallax$. The fractional
          parallax uncertainty assumed for the computation of all
          probabilities is $f=0.2$.}
        \label{fig:joint}
\end{figure}

The joint distribution $p(\parallax,\parallaxtrue)$ (i.e. the
non-normalised posterior, plotted as a 2D function of data $parallax$
and parameter $\parallaxtrue$) for this particular case of truncated
uniform stellar volume densities is depicted in
\figref{fig:joint} together with the conditional distributions for
particular values of $\parallax$ and $\parallaxtrue$. It shows
graphically the symmetry of the probability distribution
$p(\parallax|\parallaxtrue)$ (with respect to $\parallaxtrue$) and the
bias and asymmetry of $p(\parallaxtrue|\parallax)$.

\cite{LutzKelker73} obtain \equref{eq:LKposterior2} in their Sect.
     {\sc ii} under the assumption of uniform stellar volume densities
     and constant fractional parallax uncertainties (constant
     $f$). They discuss several distributions for different values of
     the ratio $\parallaxsd/\parallaxtrue$. In their Sect. {\sc iii}
     they use the expected value of the true parallax given by the
     distribution $p(\parallaxtrue|\parallax)$ in
     \equref{eq:LKposterior2} to infer the expected value of the
     difference between the true absolute magnitude $M_{True}$ and the
     value obtained with the naive inversion of the observed
     parallax. The expected value of this absolute magnitude error is
     derived and tabulated for the distribution
     $p(\parallaxtrue|\parallax)$ as a function of the fractional
     parallax uncertainty $f$. This so-called Lutz--Kelker correction
     is often applied to stellar samples that do not fulfil the
     assumptions under which it was derived because the stellar volume
     density is far from uniform at scales larger than a few tens of
     parsecs and the samples to which the correction is applied are
     never characterised by a unique value of $f$.

\subsection{Astrometry-based luminosity}

An obvious way to avoid the problems associated with the naive inversion
of observed parallaxes (see \secref{sec:Smith-Eichhorn})
is to remain in the space of parallaxes (as opposed to that of
distances) insofar as this is possible. One example of this approach
is the astrometry-based luminosity (ABL) method \citep{ABL}
originating from \cite{1920MeLuS..22....3M}. The ABL
method consists in substituting the absolute magnitudes by a proxy
that is linearly dependent on the parallax. The original proposal was

\begin{equation}
a_V\equiv 10^{0.2M_V}=\parallax 10^{\frac{m_V+5}{5}},
\end{equation}

\noindent and has been recently used to obtain maximum likelihood
estimates of the period-luminosity relation coefficients for Cepheids
and RR~Lyrae stars \citep{2017A&A...605A..79G}, and to improve the
{\it Gaia} parallax uncertainties using deconvolved colour-magnitude
diagrams as prior \citep{2017arXiv170605055A}. The new
astrometry-based luminosity depends linearly on the parallax, and thus
its uncertainty can be expected to have an approximately Gaussian
distribution if the fractional uncertainty of the apparent magnitude is
negligible. This is more often the case than for fractional parallax
uncertainties and is in general a good approximation.

Unfortunately, the astrometry-based luminosity can only be applied
 to the study of the luminosity and can do nothing for 
the analysis of spatial distributions where distances or
tangential velocities are inevitable ingredients.

%----------------------------------------------------------------
% Guide
%----------------------------------------------------------------
%----------------------------------------------------------------
% A practical guide to use DR2 parallaxes
%----------------------------------------------------------------
\section{Recommendations for using astrometric data}
  \label{sec:guide}
  
In this section we provide specific advice on the use of astrometric
data in astronomical data analysis. Although the focus is on the use
of {\it Gaia} data, many of the recommendations hold for the analysis
of astrometric data in general. To illustrate the recommendations we
also provide a small number of worked examples, ranging from very
basic demonstrations of the issues mentioned in 
\secref{sec:review} to full Bayesian analyses. Some of these examples are available in the {\it Gaia} archive tutorial described in
  \secref{sec:tutorial}.

%\textcolor{blue}{Consider stating that in the case of a single
%  parallax, with small uncertainties, inverting it may be reasonable}

\subsection{Using {\it Gaia} astrometric data: how to proceed?}

The fundamental quantity sought when measuring a stellar parallax is
the distance to the star in question. However, as discussed in the
previous sections the quantity of interest has a non-linear relation to
the measurement, $r=1/\parallaxtrue$, and is constrained to be
positive, while the measured parallax can be zero or even
negative. Starting from a measured parallax which is normally
distributed about the true parallax, this leads to a probability
density for the simple distance estimator $\rho=1/\parallax$ (see
\secref{sec:review}) for which the moments are defined
in terms of unknown quantities. This means we cannot calculate
the variance of the estimator or the size of a possible bias, which
renders the estimator useless from the statistical point of view.

{\bf Our first and main recommendation  is thus to always treat the
derivation of (astro-)physical parameters from astrometric data, in
particular when parallaxes are involved, as an inference problem which
should preferably be handled with a full Bayesian approach.}

%The discussion items below are meant to further clarify
%and qualify this recommendation.

%\subsection{Recommended methodology: full Bayesian inference of distances.}
%\label{sec:recommended}

%So far, we have discussed several approaches to infer quantities of
%astronomical interest that involve observed parallaxes. We have
%discussed the problems associated to the use of these methodologies
%and the consequences that scientists must expect from them. In the
%following we describe the method that we believe to be the best
%methodological approach to use observed parallaxes in the astronomical
%context.

%In the previous paragraphs we have described the two sources of bias
%that arise due to the distribution of sources as a function of
%parallax (the Lutz-Kelker bias) and to the naive inversion of
%parallaxes described by Eq. \ref{eq:naiveInversion}.

%The ABL method does not suffer from the caveats associated to
%the inversion of the observed parallax but it is important to 
%bear in mind that it does not account for the Lutz-Kelker bias 
%when the sample used has been truncated in distance or parallax
%(see Sec. \ref{sec:censorhips}). It only removes the bias due 
%to the non-linear transformation from parallaxes to distances or 
%absolute magnitudes but permits to work with null or negative
%parallaxes, allowing to have to avoid this kind of sample
%censorship.

\subsubsection{Bayesian inference of distances from parallaxes}
\label{sec:Bayesfromparallaxesonly}

The Bayesian approach to inference involves estimating a PDF over the
quantity of interest given the observables.  In this case we want to
estimate the distance, \dist, given the parallax, \parallax.  A fuller
treatment of this problem has been given in
\cite{2015PASP..127..994B}, so only a brief summary will be given
here.  Using Bayes' theorem we can write the posterior as

\begin{equation}
P(\dist \given \parallax) \,=\, \frac{1}{Z}P(\parallax \given \dist)P(\dist) \ .
\label{eqn:bayesdistance}
\end{equation} 

\noindent Formally, everything is also conditioned on the parallax
uncertainty, \parallaxsd, and any relevant constraints or assumptions,
but symbols for these are omitted for brevity.  The quantity
$P(\parallax \given \dist)$ is the likelihood from
\equref{eq:LKposterior}. The prior, $P(\dist)$, incorporates our
assumptions and $Z$ is a normalisation constant.

In addition to the likelihood, there are two important choices which
must be made to estimate a distance: the choice of prior and the
choice of estimator. We will first focus on the former, and start
discussing the simplest prior: the uniform unbounded prior. With a
uniform boundless (and thus improper) prior on distances the
posterior is proportional to the likelihood, and if we choose the mode
of the posterior as our estimator, then the solution is mathematically
equivalent to maximising the likelihood.  However, a boundless uniform
prior permits negative distances, which are non-physical, so we should
at least truncate it to exclude these values.

The more measurements we have, or the more precise the measurements
are, the narrower the likelihood and the lower the impact of the prior
should be on the posterior. It turns out, however,  that with the unbounded
uniform prior the posterior is improper, i.e.\ it is not 
normalisable. Consequently, the mean and median are not defined.  The
only point estimator is the mode, i.e.\ $\distest=1/\parallax$ (the
standard deviation and the other quantiles are likewise undefined),
which is rather restrictive.  Finally, this Bayesian distance estimate
defined for an unbounded uniform prior reduces to the maximum
likelihood estimate and coincides with the naive inversion discussed
in \secref{sec:parallaxinversion}. The posterior is
ill-defined for the unbounded uniform prior for parallaxes. This prior
describes an unrealistic situation where the observer is placed at the
centre of a distribution of sources that is spherically symmetric
 and the volume density of which decreases sharply
with distance.

The solution to these problems (non-physical distances, improper
posterior) is to use a more appropriate prior.  The properties of
various priors and estimators have been studied by
\cite{2015PASP..127..994B} and \cite{2016ApJ...833..119A}. The latter
makes a detailed study using a Milky Way model for a prior, and also
investigates how the estimates change when the {\it Gaia} photometric
measurements are used in addition to the parallax. One of the
least informative priors we can use is the  exponentially
decreasing space density prior:

\begin{equation}
P(\dist) \,=\,  \begin{dcases}
  \ \ \frac{1}{2\distlen^3}\,\dist^2e^{-\dist/\distlen}  & \:{\rm if}~~ r >0 \\
  \ \ 0                          & \:{\rm otherwise} \ .
\end{dcases}
\label{eqn:r2e-2prior}
\end{equation}

\noindent For distances $\dist \ll \distlen$ this corresponds to a
constant space density of stars, with the probability dropping
exponentially at distances much larger than the mode (which is at
$2\distlen$).  Examples of the shape of the posterior for parallaxes
of different precisions are shown in \cite{2015PASP..127..994B} and
\cite{2016ApJ...833..119A}.

%An alternative to using the full posterior is to collapse the
%information contained in the PDF into a single value: the point
%estimate or estimator. The mode, the mean and the median are the most
%common choices. The aforementioned series of papers examine the
%consequences of using several prior prescriptions and the
%bias/variance properties of the mode and median estimators for each of
%them.

The posterior obtained for the prior defined in 
\equref{eqn:r2e-2prior} is normalised and thus, we have a choice of point estimators (mean, median, or mode). Also, the distribution is
asymmetric, and two quantiles (5\% and 95\%) rather than the
standard deviation are recommended to summarise the uncertainty in the
point estimate. The median, as a point estimate, is guaranteed to lie
between these quantiles.  \cite{2016ApJ...833..119A} used this prior,
as well as a Milky Way prior, to infer distances for the two million
TGAS stars in the first {\it Gaia} data release. The behaviour of 
the estimates derived from the exponentially decreasing space density prior 
can be explored using the interactive tool available in the tutorial described in
\secref{sec:tutorial_comparison_estimators}.

In general, the introduction of reasonable prior probabilities
accounts for the Lutz--Kelker bias, although the inevitable mismatch
between the true distribution of parallaxes and the prior used will
result in less accurate inferences. In any case, the advantage with
respect to the methods discussed in \secref{sec:review} is clear:
{\sl i)} we do not need to tabulate corrections for each prior
assuming constant $f$; {\sl ii)} we do not need to dispose of
non-positive parallaxes; {\sl iii)} we obtain a proper full posterior
distribution with well-defined moments and credible intervals; {\sl
  iv)} even simple priors such as the exponential decreasing volume
density will improve our estimates with respect to the unrealistic
prior underlying the maximum likelihood solution
$\distest=1/\parallax$; and finally, {\sl v)} we obtain  estimators
that degrade gracefully as the data quality degrades, and with
credible intervals that grow with the observational uncertainties
until they reach the typical scales of the prior when the observations
are non-informative. These advantages come at the expense of an
inference that is more computationally demanding in general (as it
requires obtaining the posterior and its summary statistics if
needed), the need for a thoughtful choice of a prior distribution, and
the analysis of the influence of the prior on the inference results.

\begin{figure*}[htb]
        \centering
        \includegraphics[width=1.00\textwidth]{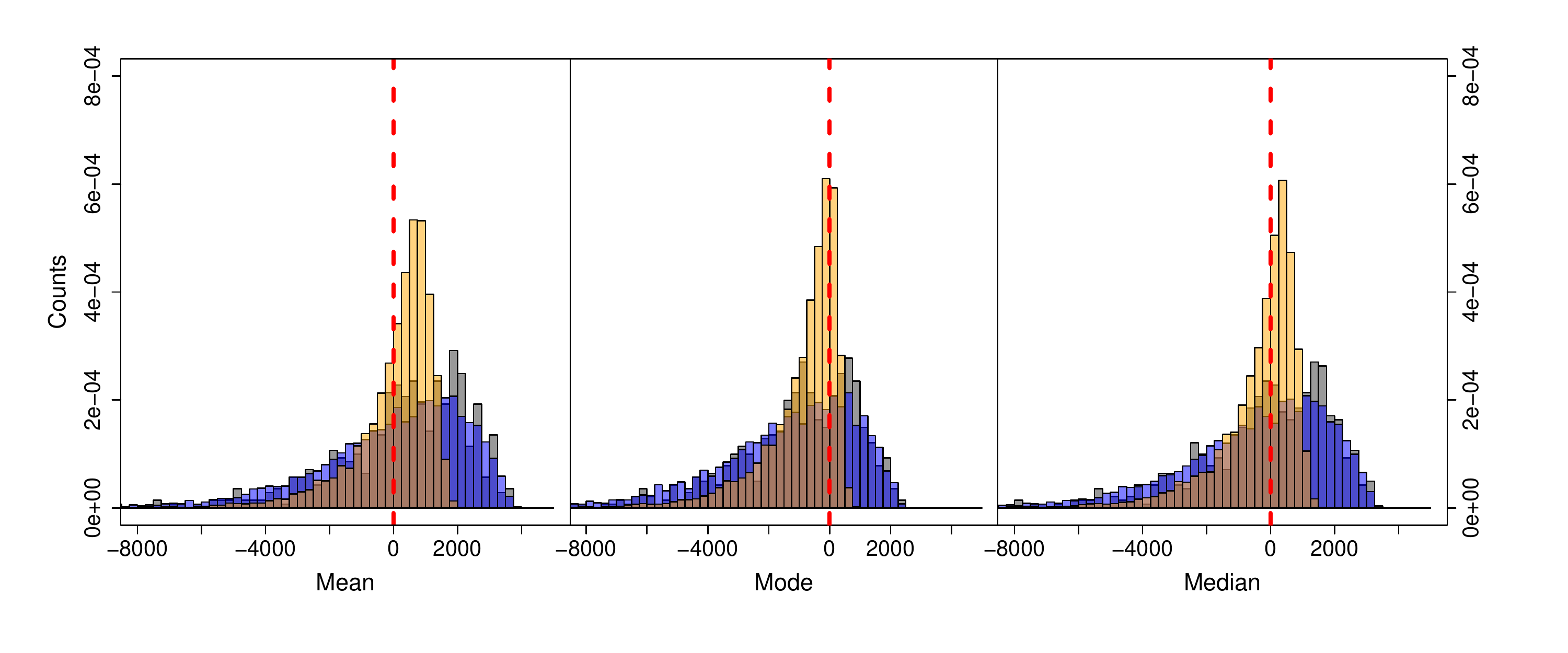}
                \caption{Probability distribution of the residuals of
                  the Bayesian estimate of the true distance in
                  parsecs for 100000 simulated stars drawn from a
                  uniform density plus exponential decay distribution,
                  and $\parallaxsd=3\cdot 10^{-1}$ mas (orange), 3 mas
                  (blue), and 30 mas (grey).}
                \label{fig:ABL-LK}
\end{figure*}

\figrefalt{fig:ABL-LK} shows the distribution of means (left), modes
(centre), and medians (right) of the posteriors inferred for a
simulation of $10^5$ sources drawn from an exponentially decreasing
space density distribution. This simulation represents the unlikely
case where the prior is a perfect representation of the true
distribution.

\begin{figure*}[htb]
        \centering
        \includegraphics[width=1.00\textwidth]{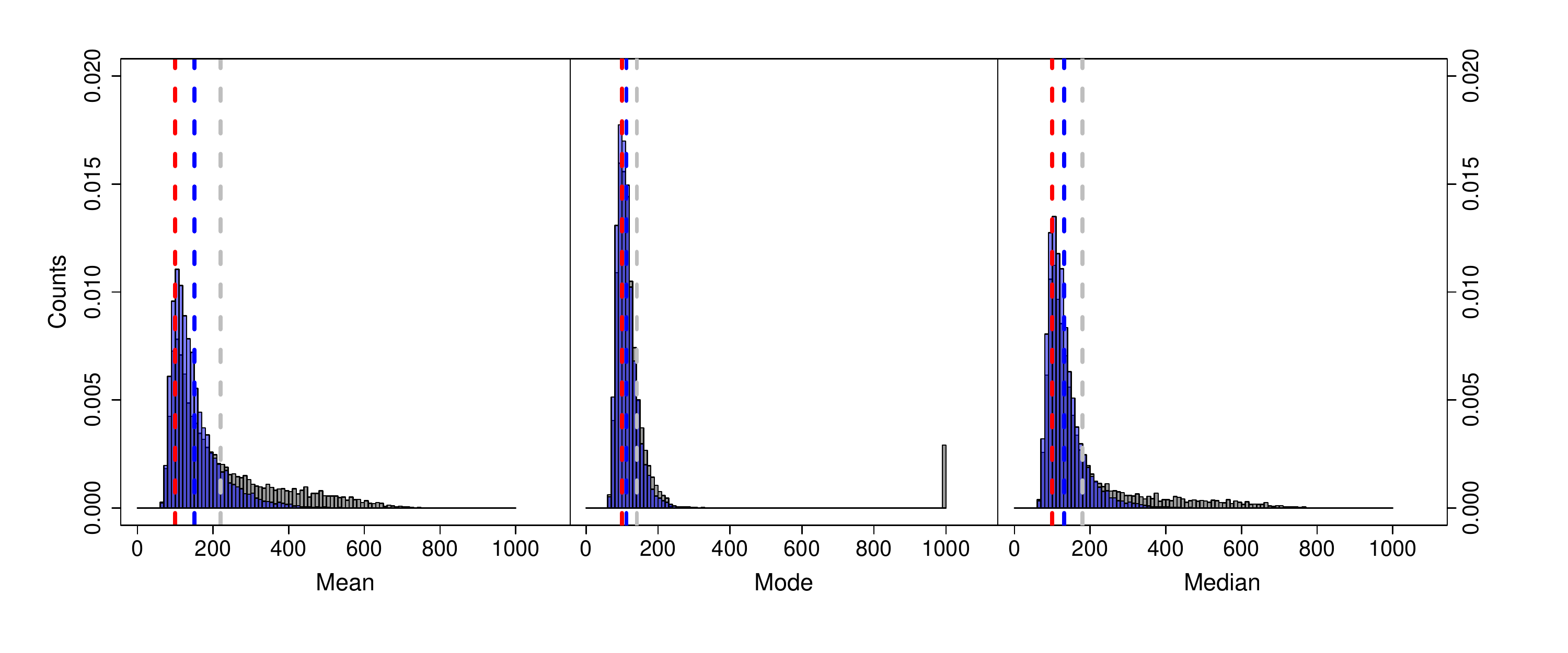}
        \caption{Distributions of means (left), modes (centre), and
          medians (right) for a series of $10^4$ posteriors calculated
          for a star at a true distance of 100 pc and $f=0.2$, and
          observed parallaxes drawn at random from the corresponding
          Gaussian distribution. Posteriors are inferred with a
          uniform density prior truncated at 1 kpc (grey) or with a
          uniform density with exponential decay prior and length
          scale 1.35 kpc (blue). The red vertical line marks the true
          parallax; the grey and blue lines correspond to the expected
          value (mean) of each distribution (same colours as 
           for the histograms).}
        \label{fig:histsmeanmodemed}
\end{figure*}

From a Bayesian perspective the full posterior PDF is the final result
of our inference if we only use parallax measurements to infer the
distance (see below), and further analyses should make use of it as a
whole. This is our recommendation in general: avoid expectations and
summaries of the posterior. However, it is often useful to compute
summary statistics such as the mean (expectation), median, mode,
quantiles, etc., to have an approximate idea of the distribution, but
we should not use these summaries for further inference, for example
to estimate absolute magnitudes, Cartesian velocities, etc. We
recommend  inferring the full posterior distributions for these derived
quantities using the posterior of the true parallax or of the
distance, or using the same Bayesian scheme as for the true parallax
as explained in \secref{sec:absmags}. In
\figref{fig:histsmeanmodemed} we show the values of the mean (left),
mode (centre), and median (right) that we would obtain from a set of
$10^4$ simulated observations of a star at 100 parsecs with
$f=0.2$. We assume a Gaussian distribution of the observations around
the true parallax. The posterior distribution is inferred using
\equref{eqn:bayesdistance} and two priors: a uniform volume density
of sources truncated at 1 kpc (results in grey) and a uniform
density of sources multiplied by an exponential decay of length scale
200 pc as defined in \equref{eqn:r2e-2prior} (in blue). The expectation
values of the histograms are shown as dashed lines of the same colour,
with the true value (100 pc) shown as a red dashed line. We see in
  general that {\it i)} the truncation has the effect of increasing
  the number of overestimated distances; {\it ii)} the three
  estimators are biased towards larger distances (smaller parallaxes),
  but the expectation of the mode is significantly closer to the true
  value; and  {\it iii)} the abrupt truncation of the prior
  results in a spurious peak of modes at the truncation distance as
  already discussed in \cite{2015PASP..127..994B}.

\begin{figure}[htb]
        \centering
        \includegraphics[width=1.00\columnwidth]{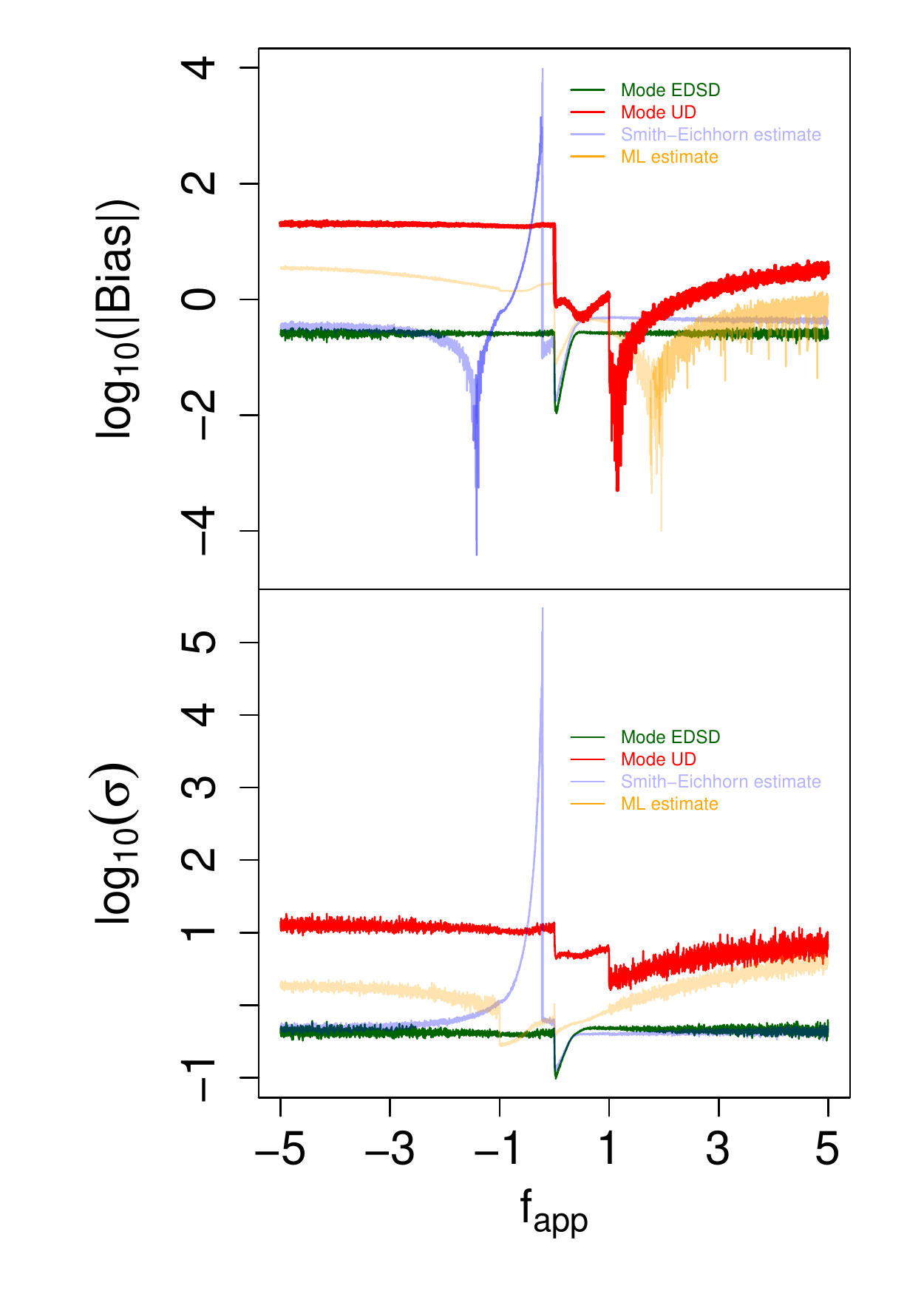}
        \caption{Bias (top) and standard deviation (bottom) averaged
          over bins of the estimated fractional parallax uncertainty
          $f_{app}$ for four estimators of the distance: the maximum
          likelihood (ML) estimator $r_{\rm est}=\frac{1}{\parallax}$
          (orange), the ML estimator corrected as described in
          \cite{SmithEichhorn} (light blue), the mode of the posterior
          obtained with an exponentially decreasing space density
          (EDSD) prior and $L=1.35$ kpc (dark green), and the mode of
          the posterior obtained with a uniform distance (UD) prior
          truncated at 100 kpc (red).}
        \label{fig:arifapp}
\end{figure}

\figrefalt{fig:arifapp} and \tabref{global} show a comparison of
the absolute value of the empirical bias and standard deviation
associated with some distance estimators discussed in this paper as a
function of the measured fractional uncertainty in the parallax. We
chose the measured value even though it is a very poor and non-robust
estimator because, as stated in \secref{sec:parallaxinversion},
we never have access to the true fractional parallax uncertainty. This
figure shows the results obtained for $10^7$ sources in the \gdrtwo
simulation described in Appendix \ref{sec:app_samples} for the {maximum likelihood} estimate $\rho=\frac{1}{\parallax}$ with and
without the Smith--Eichhorn correction, and for the mode estimates
based on the posterior distribution for two priors (a uniform distance
prior,  UD, with maximum distance $r_{lim}=100$ kpc, and an exponentially decreasing space density prior, 
EDSD, with $L=1.35$ kpc), neither of which matches the true distribution of
sources in the simulation. Only the mode of the posteriors is plotted
(but not the mean or the median) for the sake of clarity. The
conclusions described next are only valid under the conditions of the
exercise and are provided as a demonstration of the caveats and
problems described in previous sections, not as a recommendation of
the mode of the posterior inferred under the EDSD prior as an
estimator. At the risk of repeating ourselves, we emphasise the need
to adopt priors adapted to the inference problem at hand. Also, the
conclusions only hold  for the used simulation (where we generate the
true distances and hence can calculate the bias and standard deviation)
and need not be representative of the true performance for the real
\gaia data set. They can be summarised as follows:
\begin{itemize}
\item the mode of the EDSD prior shows the smallest bias and standard
  deviation in practically the entire range of estimated fractional
  parallax uncertainties (in particular, everywhere beyond the range
  of $f_{\rm app}$ represented in the plot);
\item the Smith--Eichhorn estimate shows pathological biases
  and standard deviations in the vicinity of the supposedly best-quality measurements at $f_{app}=0$. Away from this region, it
  provides the next less biased estimates (averaged over bins of
  $f_{app}$) after the mode of the EDSD posterior;
\end{itemize}

\begin{table*}[hhh]
\caption{Average bias and standard deviation in three regimes of
  $f_{\rm app}$ for four distance estimators discussed in this
  paper: (from left to right) the mode of the posterior based on the
  exponentially decreasing space density (EDSD) prior; the mode of the
  posterior of the uniform distance (UD) distribution; the maximum
  likelihood estimate corrected according to \cite{SmithEichhorn}
  abbreviated as SE; and the maximum likelihood (ML) estimate. The
  wider ranges of $f_{\rm app}$ exclude narrower ranges shown in
  previous rows of the table.}  \centering
\label{global}
\begin{tabular}{llcccc}
\hline \hline
Summary & $f_{\rm app}$ Range & EDSD & UD & SE & ML \\
\hline
\multirow{3}{*}{Bias} &(-1,1)  & -0.2  &9.7  & 34.2 &-0.95 \\ 
                      &(-5,5)  & -0.3  &10.7 &-0.34 &-1.2 \\
                      &(-50,50)& -0.3  &16.2 &-0.4  &-3.8 \\
\hline
\multirow{3}{*}{Std. Deviation} &(-1,1)  &  0.4    &8.0   &685.8 &   0.5 \\
                      &(-5,5)  &  0.4    &8.4   &  0.5 &   1.95 \\
                      &(-50,50)&  0.4    &10.6  &  0.4 &  17.1  \\
\hline
\end{tabular}
\end{table*}

\subsubsection{Bayesian inference of distances from parallaxes and
  complementary information}
\label{sec:hierarchical-Bayes}

The methodology recommended in the previous paragraphs is useful when
we only have observed parallaxes to infer distances. There are,
however, common situations in astronomy where the parallaxes are only
one of many observables, and distances are not the final goal of
the inference problem but a means to achieve it. In this context, we
recommend an extension of the classical Bayesian inference methods
described in the previous section. These problems are characterised by
a set of observables and associated uncertainties (that include but
are not restricted to parallaxes) and a series of parameters (the
values of which are unknown a priori) with complex interdependence
relationships amongst them. Some of these parameters will be the
ultimate goal of the inference process.  Other parameters do play an
important role, but we are not interested in their particular
values, and we  call them {\sl nuisance} parameters, following
the literature. For example, in determining the shape of a stellar
association, the individual stellar distances are not relevant by
themselves, but only insomuch as we need them to achieve our
objective. We  show below how we deal with the nuisance parameters.
The interested reader can find applications of the methodology
described in this section to inferring the coefficients of
period-luminosity relations in \cite{2017A&A...605A..79G} and
\cite{2017ApJ...838..107S}. Also, the same methodology (a hierarchical
Bayesian model) is applied in \cite{2017MNRAS.471..722H} where the
constraint on the distances comes not from a period-luminosity
relation, but from the relatively small dispersion of the absolute
magnitudes and colour indices of red clump stars. A last example of
this methodology can be found in \cite{2017arXiv170308112L} where the
constraint comes from modelling the colour-magnitude diagram.

Just as in the previous section where we aimed at estimating distances
from parallaxes alone, the two key elements in this case are the
definitions of a likelihood and a prior. The likelihood represents the
probability distribution of the observables given the model
parameters. Typically, the likelihood is based on a {\em
generative} or {\em forward} model for the data. Such models predict
the data from our assumptions about the physical process that
generates the true values (i.e. the distribution of stars in space)
and our knowledge of the measurement process (e.g. justifying the
assumption of a normal distribution of the observed parallax around its
true value). Forward models can be used to generate arbitrarily large
synthetic data sets for a given set of the parameters. In this case,
however, where we are concerned with several types of measurements
that depend on parameters other than the distance, the likelihood
term will be in general more complex than in
\secref{sec:Bayesfromparallaxesonly} and may include probabilistic
dependencies between the parameters. The term {\sl hierarchical} or
{\sl multi-level} model is often used to refer to this kind of model.

Let us illustrate the concept of hierarchical models with a simple
extension of the Bayesian model described in
\secref{sec:Bayesfromparallaxesonly}, where instead of assuming a
fixed value of the prior length scale $L$ in \equref{eqn:r2e-2prior},
we make it another parameter of the model and try to infer it. Let us
further assume that we have a set of $N$ parallax measurements
$\{\parallax_k\}$, one for each of a sample of $N$ stars. In this
case, the likelihood can be written as
%
%\begin{equation}
%p(\{\parallax_k\}|\{r_k\},L) = p(\{\parallax_k\}\mid \{r_k\})\cdot
%  p(\{r_k\}\mid L)=p(\{\parallax_k\}\mid \{r_k\})\cdot \prod_{k=1}^N%  p(r_k\mid L). 
%\end{equation}
\begin{eqnarray}
p(\{\parallax_k\}|\{r_k\},L) & = & p(\{\parallax_k\}\mid \{r_k\})\cdot p(\{r_k\}\mid L)  \nonumber \\
  & = & p(\{\parallax_k\}\mid \{r_k\})\cdot \prod_{k=1}^N p(r_k\mid L), 
\end{eqnarray}
where $r_k$ is the true unknown distance to the $k$th star.  We note that
very often the measured parallaxes are assumed independent, and thus
$p(\{\parallax_k\}\mid \{r_k\})$ is written as the product
$\prod_{k=1}^N p(\parallax_k \mid r_k)$. This is incorrect in general for
{\gaia} parallaxes because the parallax measurements are not
independent. As described in \citet{DR2-DPACP-51} and
\secref{section:gaiadata} of this paper, there are regional
correlations amongst them (see \secref{sec:summary_recommendations}), 
but for the sake of simplicity let us assume the sample of $N$
measurements is spread all over the celestial sphere such that the
correlations average out and can be neglected. Hence, we write

\begin{equation}
p(\{\parallax_k\}|\{r_k\},L) = \prod_{k=1}^N p(\parallax_k \mid r_k)\cdot 
p(r_k\mid L).
\label{eq:hbm-ex1}
\end{equation}

\noindent Under the assumption of Gaussian uncertainties, the first term in the
product is given by \equref{eq:pdf_truepi}, while the second is given
by \equref{eqn:r2e-2prior}.

\begin{figure}
\begin{tikzpicture}
\tikzset{vertex/.style = {shape=circle,draw,minimum size=40pt}}
\tikzset{edge/.style = {->,> = latex',very thick}}
% vertices
\node[vertex] (a) at  (2,8) {$L$};
\node[vertex] (b) at  (2,5) {$r_k$};
\node[vertex,fill=gray] (c) at  (2,2) {$\parallax$};
%edges
\draw[edge] (a) to (b);
\draw[edge] (b) to (c);
\draw (0,0) rectangle (4,9);

\draw (5,2) coordinate (M);
\filldraw[black] (M) circle (2pt);
\node (repeat) at (5.25,2.0) [right] {$\parallaxsd$}; 
\draw[edge] (M) to (c);

\draw (5,8.5) coordinate (N);
\filldraw[black] (N) circle (2pt);
\node (repeat) at (5.25,8.5) [right] {$\mu_L$}; 
\draw[edge] (N) to (a);

\draw (5,7.5) coordinate (O);
\filldraw[black] (O) circle (2pt);
\node (repeat) at (5.25,7.5) [right] {$\sigma_{L}$}; 
\draw[edge] (O) to (a);

\node (repeat) at (1,0.5) [right] {$k=1,2,...,N$}; 

\end{tikzpicture}
\caption{Directed acyclic graph that represents a hierarchical
  Bayesian model of a set of $N$ parallax measurements characterised
  by uncertainties $\parallaxsd$ and true distances drawn from an
  exponentially decreasing density distribution of distances (see
  \equref{eqn:r2e-2prior}). The scale length of the exponential
  decrease, $L$, is itself a model parameter that we can infer from
  the sample. Its prior is defined in this case for the sake of
  simplicity as a Gaussian distribution of mean $\mu_L$ and standard
  deviation $\sigma_L$.}
\label{graph-ex1}
\end{figure}
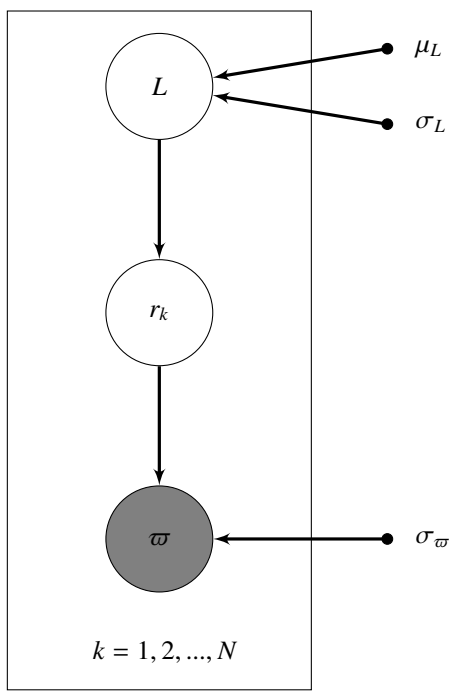

This likelihood can be represented by a simple directed graph (see
\figref{graph-ex1}) that provides information about the conditional
dependencies amongst the parameters. The shaded nodes represent the
observations, the open circles represent model parameters, and the
small black circles at the origin of the arrows represent model
constants. The arrows denote conditional dependence relations, and the
plate notation indicates repetition over the measurements $k$.

The next key element is, as in \secref{sec:Bayesfromparallaxesonly},
the prior.  According to \figref{graph-ex1}, the only parameter
that needs a prior specification is the one without a parent node:
$L$. The rest of the arcs in the graph are defined in the likelihood term
(\equref{eq:hbm-ex1}).  If the sample of $N$ stars were
representative of the inner Galactic halo for example, we could use a
Gaussian prior centred at $\approx 30$ kpc \cite[see e.g.][and
  references therein]{2018MNRAS.474.2142I}.  Such a hierarchical model
can potentially shrink the individual parallax uncertainties by
incorporating the constraint on the distribution of distances.

If we are only interested in the individual distances $r_k$, we can
consider $L$ as a nuisance parameter:

%\begin{equation}
%p({\parallaxtrue}_{;k} \mid \{\parallax_{k}\}) = \int
%p({\parallaxtrue}_{;k},L \mid \{\parallax_{k}\})\cdot {\rm d}L= \int
%p({\parallaxtrue}_{;k} \mid \{\parallax_{k}\},L)\cdot
%p(L\mid\{\parallax_{k}\})\cdot {\rm d}L.
%\label{eq:marginalisation}
%\end{equation}

\begin{eqnarray}
p({\parallaxtrue}_{;k} \mid \{\parallax_{k}\}) & = & \int p({\parallaxtrue}_{;k},L \mid \{\parallax_{k}\})\cdot {\rm d}L  \label{eq:marginalisation} \\
  & = & \int p({\parallaxtrue}_{;k} \mid \{\parallax_{k}\},L)\cdot p(L\mid\{\parallax_{k}\})\cdot {\rm d}L. \nonumber
\end{eqnarray}

\noindent This integral (known as the marginalisation integral)
  allows us to write the posterior we are interested in without having to
  fix the value of $L$ to any particular value. Depending on the
  objective of the inference, we could have alternatively determined
  the posterior distribution of $L$ by marginalising the individual
  distances with an $N$-dimensional integral over the $\{r_k\}$
  parameters.

In parameter spaces of dimensionality greater than 3--4 the computation
of the possibly marginalised posteriors and/or evidence requires 
efficient sampling schemes like those inspired in Markov chain Monte
Carlo (MCMC) methods to avoid large numbers of calculations in regions
of parameter space with negligible contributions to the
posterior. This adds to the higher computational burden
of the Bayesian inference method mentioned in the previous section.

The previous simple example can be extended to include more levels in
the hierarchy and, more importantly, more parameters and measurement
types. \secrefalt{subsec:lumin-cal} and \ref{subsec:PLrelation}
develop in greater detail two examples of hierarchical models of direct
applicability in the \gaia context.

%%%%%%%%%%%%%%%%%%%%%%%%%%%%%%%%%%%%%%%%%%%%%%%%%%%%%%%%%%%%%%%%%%%%%%%%%
\subsection{Absolute magnitudes, tangential velocities, and other derived quantities.}
\label{sec:absmags}

The approaches described in the previous sections can be applied to
any quantity  we want to estimate using the parallax. For
example, if we want to infer the absolute magnitude \mg, then given
the measured apparent magnitude \gmag\ and line-of-sight extinction
\ag, the true parallax \parallaxtrue\ is related to \mg\ via the
conservation of flux

\begin{equation}
5\log\parallaxtrue \,=\, \mg + \ag - \gmag - 5  \ .
\label{eqn:mg}
\end{equation}

Assuming for simplicity that \gmag\ and \ag\ are known, Bayes' theorem
gives the posterior on \mg\ as

\begin{equation}
P(\mg \given \parallax, \gmag, \ag) \,=\ \frac{1}{Z}P(\parallax \given \mg, \ag, \gmag)P(\mg)
\label{eqn:bayes_mg}
,\end{equation}

% formally there is a term P(\gmag, \ag | \mg) on the RHS, but for
% fixed \gmag and \ag this is a delta function.
\noindent where the likelihood is still the usual Gaussian
distribution for the parallax (\equref{eq:pdf_truepi}) in which the
true parallax is given by \equref{eqn:mg}. As this expression is
non-linear, we  again obtain an asymmetric posterior PDF over \mg,
the exact shape of which also depends  on the prior.

The inference of other quantities can be approached in the same way. In
general we  need to consider a multi-dimensional likelihood to
accommodate the measurement uncertainties (and correlations) in all
observed quantities. For instance, the set of parameters
  ${\boldsymbol \theta}=\{r,v,\phi\}$ (distance, tangential speed, and
  direction measured anticlockwise from north) can be inferred from
  the \gaia astrometric measurements
  $\mathbf{o}=\{\parallax,\mu_{\alpha^*},\mu_{\delta}\}$ (where
  $\mu_{\alpha^*}$ and $\mu_{\delta}$ are the measured proper
  motions) using the  likelihood

\begin{equation}
  p(\mathbf{o}\mid \boldsymbol{\theta})=
  \mathcal{N}(\boldsymbol{\theta},\Sigma)=\frac{1}{(2\pi)^{3/2}|\Sigma|^{1/2}}
  \exp\left(-\frac{1}{2}(\mathbf{o}-\mathbf{x})^T \Sigma^{-1}
  (\mathbf{o}-\mathbf{x})\right),
\label{eq:tanvel}
\end{equation}

\noindent where $\mathcal{N}$ denotes the Gaussian distribution,
  $\Sigma$ is the full (non-diagonal) covariance matrix provided as
  part of the \gaia Data Release, and 

\begin{equation}
\mathbf{x} = \left(\frac{1}{r},\frac{v\sin(\phi)}{r},\frac{v\cos(\phi)}{r}\right)
\end{equation}

\noindent is the vector of model parameters geometrically
  transformed into the space of observables in noise-free
  conditions. \equrefalt{eq:tanvel} assumes correlated Gaussian
  uncertainties in the measurements . 

  The posterior distribution can then be obtained by multiplying
  the likelihood by a suitable prior. The simplest assumption would be
  a separable prior such that $p(\boldsymbol\theta)=p(r)\cdot
  p(v)\cdot p(\phi)$, where $p(v)$ and $p(\phi)$ should reflect our
  knowledge about the dynamical properties of the population from
  where the source or sources were drawn (e.g.\ thin disk, thick disk,
  bulge, halo). Again, hierarchical models can be used in the
  analysis of samples in order to infer the population properties
  (prior hyper-parameters) themselves.

  Similar procedures can be followed to infer kinematic
  energies or full 3D velocities when the forward model is extended
  with radial velocity measurements.

\subsection{Further recommendations}
\label{sec:summary_recommendations}

In this subsection we provide some further recommendations and
guidance in the context of the Bayesian approach outlined above.
Although powerful, inference with Bayesian methods usually comes at a
large computational cost. Some of the recommendations below can also
be seen in the light of taking data analysis approaches that
approximate the Bayesian methodology and can be much faster.

\paragraph{\bf Where possible, formulate the problem in the data
space} The problems caused by the ill-defined uncertainties on
quantities derived from parallaxes can be avoided by carrying out the
analysis in the data space where the behaviour of the uncertainties
is well understood. This means that the quantities to be inferred are
treated as parameters in a forward or generative model that is used to
predict the data. Some adjustment process then leads to estimates of
the parameters. A very simple forward modelling example can be found
in \cite{2004A&A...428..149S} who studied the luminosity calibrations
of O-stars by predicting the expected {\hip} parallaxes from the
assumed luminosity calibration and comparing those to the measured
parallaxes.  A more complex example can be found in
\cite{2000A&A...356.1119L} who present a kinematic model for clusters
which describes the velocity field of the cluster members and predicts
the proper motions, accounting for the astrometric uncertainties and
covariances. As shown in previous sections, the Bayesian approach
naturally lends itself to (and in fact requires) forward modelling of
the data. 

Forward modelling has the added advantage that it forces us to
consider the proper formulation of the questions asked from the
astrometric data. This naturally leads to the insight that often the
explicit knowledge of the distances to sources is not of interest. For
example, in the \cite{2004A&A...428..149S} case an assumed luminosity
of the O-stars and their known apparent magnitude is sufficient  to predict
the observed parallaxes.  In more complex analyses the distances to
sources can often be treated as nuisance parameters, which in a
Bayesian setting can be marginalised out of the posterior.

\paragraph{\bf Use all relevant information}
Although the parallax has a direct relation to the distance of a star,
it is not the only measurement that contains distance information. The
apparent magnitude and the colour of a star carry significant
information on the plausible distances at which it can be located as
the colour provides information on plausible absolute magnitude values
for the star.  This  is used in two papers
\citep{2017AJ....154..222L, 2017arXiv170605055A} in which the
information contained in the photometry of stars is combined with
{\gdrone} parallax information to arrive at more precise
representations of the colour-magnitude diagram than can be obtained
through the parallaxes alone. Similarly, proper motions contain
distance information which can be used to good effect to derive more
precise distances or luminosities for open cluster members
\citep[e.g.][]{2001A&A...367..111D, 2017A&A...601A..19G}. The Bayesian approach naturally
allows for the combination of different types of data in the modelling
of the problem at hand. It should  be emphasised, however,  that adding
additional data necessitates increasing the model complexity (e.g. to include the relation among apparent magnitude, absolute
magnitude, extinction, and parallax) which leads to the need to make
more assumptions.

\paragraph{\bf Incorporate a proper prior}
\cite{2015PASP..127..994B} discussed the  simplest case of inferring
the distance to a source from a single observed parallax. He showed
that if  the minimal prior that $r$ should be positive is used,  the
resulting posterior is not normalisable and hence has no mean,
variance, other moments, or percentiles. This behaviour of the
posterior is not limited to inference of distance. Examples of other
quantities that are non-linearly related to parallax are absolute
magnitude ($M\propto\log_{10}\varpi$), tangential velocity
($v_\mathrm{T}\propto1/\parallax$), kinetic energy, and angular
momentum (both proportional to $1/\parallax^2$ when determined
relative to the observer). In all these cases it can be shown that the
posterior for an improper uniform prior is not normalisable and has no
moments. The conclusion is that a proper prior on the parameters to be
estimated should be included to ensure that the posterior represents a
normalised probability distribution. In general using unconstrained
non-informative priors in Bayesian data analysis (such as the one on
$r$ above) is bad practice. Inevitably, there is always a mismatch
between the prior and the true distribution (if there were not, there
would be no need to do the inference). This will unavoidably lead to
some biases, although the better the data, the smaller these will be.
We cannot expect to do much better than  our prior knowledge in case
we only have access to poor data. The Bayesian approach guarantees a
graceful transition of the posterior into the prior as the data
quality degrades.

The above discussion raises the question of what priors to include for
a specific inference problem.  Simple recipes cannot be given as the
prior to be used depends on the problem at hand and the information
already available.  When deciding on a prior we should keep in mind
that some information is always available. Even if  a parallax is only
available for a single star, we know that it cannot be located at
arbitrary distances. It is unlikely to be much closer that 1 pc
(although we cannot fully exclude the presence of faint stars closer
than Proxima Centauri) and it must be located at a finite distance (we
can observe the star). This would suggest a non-informative uniform
prior on $r$ with liberal lower and upper bounds (such that the prior
is normalised).  However, as pointed out in \cite{2015PASP..127..994B}
a uniform distribution in $r$ implies a space density of stars that
falls of as $1/r^2$ from the position of the Sun, which is of course
physically implausible. Hence one should assume a reasonable
distribution of stars in space and construct the prior on $r$
accordingly. \cite{2015PASP..127..994B} presents a prior derived from
a uniform space density with an exponential cut-off which was used in
\cite{2016ApJ...833..119A} to derive distances to stars for which
parallaxes are listed in {\gdrone}. This prior should not be used
indiscriminately, at the very least we should carefully consider the
choice of the scale length $L$ (or leave that as a parameter to be
estimated, as described in \secref{sec:hierarchical-Bayes}) and in
most cases a more tailored prior based on our broad knowledge of the
distribution of a given stellar population in the Milky Way would be
better. The tutorial cases introduced in the next section contain some
more examples of priors on distance and other astrophysical
parameters.

The next two items discuss simplifications to the Bayesian
approach that nevertheless need to be justified carefully.

\paragraph{\bf Maximum likelihood modelling} We have seen that priors
are the bridges that allow us to go from the probability of the
observations given the unknown parameters to the desired
probability of the parameters given the observations. It is only based
on this probability distribution that we can make statements
about credible intervals (uncertainties) for the inferred quantities,
or select amongst competing models (a topic that is beyond the scope
of this paper). However, if making prior-free inferences is preferred,
then maximising the likelihood is the only alternative. The kinematic
modelling presented in \cite{2000A&A...356.1119L} is a non-trivial
example of this. A more complex example can be found in
\cite{2014A&A...564A..49P}. The ML approach, just as the Bayesian
framework described above, allows  the combination of different
types of data, and accounts for selection functions or missing data.
We have seen in \secref{sec:Bayesfromparallaxesonly} that the maximum
likelihood estimate of the distance given a single parallax
measurement is $\rho=\frac{1}{\parallax}$ and this is a poor estimator
of the distance except for subsets of very accurate measurements. In
general, the Bayesian and the maximum likelihood estimates coincide in
the limit of very small uncertainties or infinite numbers of
measurements. In such limits, the maximum likelihood estimate is
simpler to obtain,  although its computational cost may still be large
as the ML method is often equivalent to a complex optimisation problem
involving a multi-dimensional function of many parameters.

\paragraph{\bf Selecting the `best' data}
Analyses that use parallax data are often restricted to positive
parallaxes with relative uncertainties below some limit (typically 20
\%). This allows working in a regime where the uncertainties
of derived quantities such as distance, tangential velocities,
luminosity, etc.,\ are thought to be manageable, which allows working
in the space of astrophysical variables rather than the data.
Truncation on relative parallax error might be justified in an
exploratory phase of the data analysis; however, there are a number of
reasons why this approach is almost never advisable.  Even at relative
uncertainties below $0.2$ the quantities calculated from the parallax
can be biased and suffer from a large variance
\citep[see][]{2015PASP..127..994B}. More importantly, however, the
selection of `good' parallaxes will bias the sample studied to nearby
and/or bright representatives of any stellar population, and the
selection may lead to discarding a very large fraction of the
potential sample. Hence any inferences based on such data will be
severely biased.  This is clearly illustrated in
\figref{fig:Hist_R_bias_50percent_pi} where for an even less strict
truncation of stars with a relative uncertainties below 50\%
the distribution of distances of the resulting sample is clearly
biased with respect the original sample.

\paragraph{\bf Accounting for data selection and incompleteness}
Although the {\it Gaia} survey is designed such that the only
selection of sources is that they are brighter than the survey limit
at $G=20.7$, the combination of the onboard detection algorithm, the
{\it Gaia} scanning law, and the filtering of results of insufficient
quality prior to a data release, lead to a complex selection
function, especially in the early data releases. This selection function 
should be taken into account in any data analysis and this is most naturally 
done as part of a Bayesian
analysis where the selection function is part of the forward model
that predicts the data given the model parameters. Precise prescriptions of the selection functions are not foreseen to
be part of the data release documentation.  Hence, selection function
parameters need to be included as part of the parameters inferred by
the Bayesian analysis or, if this is not possible, the selection
functions have to be borne in mind when interpreting the results.

\paragraph{\bf Covariances in the uncertainties}
All the uncertainties on the astrometric data quoted in the {\it Gaia}
catalogue are presented as full covariance matrices, where the
diagonal elements represent the standard uncertainties on the astrometric parameters, and the off-diagonal elements the covariances or
correlations between the uncertainties. This amounts to treating the
astrometric data vector as having been drawn from a multivariate
normal distribution with a covariance matrix as given in the {\it
  Gaia} catalogue. The covariances are most easily handled in the data
space as part of the likelihood \citep[see][for an example in the
  context of kinematic modelling]{2000A&A...356.1119L}. If the
covariances in the astrometric uncertainties are not accounted for, we
can easily be misled, for example, by spurious features in the
inferred velocity field of an open cluster
\citep{1997ESASP.402...63B}.

The uncertainties are also correlated from one source to the next,
especially over small distances on the sky. A study of the
star-to-star correlations in the parallax uncertainties in {\gdrone} was done
for the Kepler field \citep{2017ApJ...844..166Z} where independent and
precise asteroseismic distances to the stars are available, enabling
the authors to derive an expression for the correlation strength and
spatial scale. This expression can be used for studies of the Kepler
field, but care should be taken when extrapolating to other fields on
the sky. The functional form for the star-to-star correlations used by
\cite{2017ApJ...844..166Z} could be introduced as part of the forward
model, with the \cite{2017ApJ...844..166Z} parameters as a good first
guess.

For {\gdrone} the length scale for the star-to-star correlations was
estimated to  vary from subdegree scales to tens of degrees on the
sky, where \cite{2017ApJ...844..166Z} derived the correlation
function over length scales of $\sim0.2$ to $\sim10$ degrees. For
{\gdrtwo} \cite{DR2-DPACP-51} estimate that the spatial correlations
extend over scales of below 1 degree to 10--20 degrees.

\paragraph{\bf Accounting for non-Gaussian and/or systematic uncertainties}
Although the bulk of the sources in the {\it Gaia} catalogue have
normally distributed uncertainties, there is a significant fraction
for which the uncertainties exhibit non-Gaussian behaviour (e.g. when uncertainties are over- or underestimated). This can be
accounted for in the data analysis by including the uncertainties as
part of the forward model \citep[e.g.][]{2017ApJ...838..107S} or by
explicitly modelling an outlier population. Similarly, systematic
uncertainties can be included as part of the forward model. For example,
\cite{2017ApJ...838..107S} include a global parallax zero-point as
part of their probabilistic model used to analyse the
period-luminosity relation of RR Lyrae stars with {\gdrone} data.
An alternative approach to the investigation of systematics in the parallaxes (or distance
  estimates obtained from\ photometry or spectroscopy, for example) is presented in
  \cite{2012MNRAS.420.1281S} and is applied to {\it Gaia} DR1 in \cite{2017MNRAS.472.3979S}. In this
  case we can consider that for samples covering a significant fraction of the sky, any
  systematic error in the estimated distances to the stars will show up as correlations in their
  3D velocity components. The presence of such correlations can be used to make
inferences about systematic errors, for example, in the parallaxes.

Systematic uncertainties are more difficult to handle as they may show variations as a function of source brightness or celestial
position, they may be correlated across neighbouring sources, and they
may not be well understood for a given early {\it Gaia} data
release.
In general the information needed to accurately model systematic
uncertainties or uncertainty correlations between sources may not be
readily available. This information can be obtained from a comparison
between {\gaia} and other high-precision data
\citep[e.g.][]{2017ApJ...844..166Z, Arenou_2017, DR2-DPACP-39} or by
examining, for example, plots of the parallax or proper motion
averaged over sky regions for samples where the true parallax or proper
motion values can be assumed to be known,  such as zero parallax and
proper motion for quasars (see \citealt{DR2-DPACP-51} for examples).

Two special cases should be mentioned: when the sample is well distributed over
the sky, we can  safely assume that the local systematics vanish and
that only the global parallax zero-point need to be subtracted; locally,
we may be interested not by the absolute value of the parallaxes, but by 
the relative ones, in which case the difference between parallaxes
and their average removes part of the systematics.

There is no general recipe for dealing with non-Gaussian uncertainties
or correlated systematic uncertainties. The main advice we can give
here is to proceed with the analysis of the astrometric data as they
are, but to keep in mind the systematics and correlations discussed in
\secref{section:gaiadata} when interpreting the results. Alternatively,
 the forward model can be extended to include systematic and correlation
terms for which parameters are also to be estimated. Such models can
be guided by the studies of systematic uncertainties mentioned above.

\paragraph{\bf Testing with simulations}
Finally, we strongly advise that the inference problem at
hand should be investigated through simulated data, and that the
simulations performed should be  as close as possible to the real data (in particular
correctly modelling the uncertainties and selection effects). The
simulations allow  the analysis method to be developed  and tested for
accuracy.  However,  the performance of the
analysis method should be interpreted strictly in terms of how well the assumed model
explains the simulated observed data. That is, we should not fall into
the trap of trying to tune the analysis method to get an answer that
is as close to the `truth' as possible. In real problems we can only
judge the adequacy of a model and its parameter values by how well
they predict the observed data (to within the observational
uncertainties, it should be stressed, as we should avoid
`over-fitting' the data).

%----------------------------------------------------------------
% Tutorials
%----------------------------------------------------------------
%----------------------------------------------------------------
% A practical guide to use DR2 parallaxes
%----------------------------------------------------------------
\section{Using astrometric data: practical examples}
  \label{sec:tutorial}
  
%{\bf \large This section is closely linked to the tutorials (in R and Python) provided
%in the {\it Gaia} archive pages. They can be found in this link: \textcolor{blue}{URL}}

We introduce here a few worked examples to illustrate the points that were made in the previous
section. These examples are available in full detail as online tutorials in the form of source code,
accompanied by much more extensive explanation than can be provided here. The source code and
corresponding Python and R Notebooks can be found at the following URL: \toprepo. In all cases the
reader is strongly encouraged to download the source code and experiment with modifications of the
simulated input data and/or the parameter choices in the inference methods.

\subsection{Comparison of distance and distance modulus estimators}
\label{sec:tutorial_comparison_estimators}

The use of the Bayesian inference with non-informative priors described
in \secref{sec:Bayesfromparallaxesonly} is illustrated and implemented
in the following tutorial \subrepo{single-source/GraphicalUserInterface/}. The tutorial
 compares the performance of Bayesian distance estimation methods 
with the Smith--Eichhorn transformation \citep{SmithEichhorn} (\secref{sec:Smith-Eichhorn}) 
and the naive parallax inversion. 

The tutorial contains a Graphical User Interface that  easily visualises 
and compares the behaviour of all these estimators for a given parallax and
uncertainty. For the Bayesian inference, estimations using the mode and the median
are provided together with a 90\% confidence interval. The tutorial 
also provides a library, {\it pyrallaxes}, with the implementation of all these estimators.
The library can easily be customised to implement other priors for the Bayesian inference.

Additionally, an implementation of the Bayesian distance estimator using the {\it Exponentially Decreasing Space Density} prior
introduced in \cite{2015PASP..127..994B} will be available in TopCat (\url{http://www.starlink.ac.uk/topcat/}) and Stilts (\url{http://www.starlink.ac.uk/stilts/}) from respectively versions 4.6 and 3.1-3 onwards.

\subsection{Inferring the distance to a single source using just the parallax}

The issues surrounding the use of a parallax to infer a distance were explored in
\cite{2015PASP..127..994B} and applied to simulated \gaia data in \cite{2016ApJ...832..137A} and to
TGAS (\gdrone) in \cite{2016ApJ...833..119A}. A tutorial exploring this is provided at
\subrepo{single-source/tutorial}. This can be used to investigate how the posterior depends on the
prior and the data. It also includes a simple example of a hierarchical model to avoid specifying
the exact length scale of a distance prior.

\subsection{Inferring the distance to and size of a cluster using just the parallaxes}

In many applications we are more interested in the average distance to a cluster (or a group of
stars) rather than to the individual stars. In this case a basic mistake to be avoided is
estimating the individual distances (whatever the method) and then averaging these
individual values. A more correct approach is to average the individual parallaxes and
then to obtain a distance from this average value. However, a much better solution is to set up a model for the cluster,  which
would use as parameters the distance to its centre, for example, and some measure of its size,  and to infer
its parameters. This is explored in the tutorial at \subrepo{multiple-source/}. This introduces the
overall problem and derives a general solution. Code is implemented for the specific case of a model
which assumes a random isotropic distribution of the true stars from the centre of the cluster. This
model has two parameters, the cluster distance and the cluster size. The solution uses a small angle
approximation to make the problem simpler, although it is easily extended to the case of clusters
with a significant angular extent. It is applied to the Pleiades selection from the {\gdrone} main
release paper \citep{2016A&A...595A...2G}. The tutorial also considers the problem of how to
accommodate correlations in the measured parallaxes of different stars. Finally, it also shows the
results from a classical and  a naive combination of stellar parallaxes to estimate the cluster
distance. The combination of parallaxes and proper motions of individual stars in a cluster into a
single solution for the mean parallax and proper motion is treated as an iterative least squares
problem in \cite{2017A&A...601A..19G} (see their Appendix A for details).

\subsection{Inferring the distance and velocity of a source using the parallax and proper motions}

The velocity (speed and direction) of a source in the plane of the sky can be inferred from
measurements of its parallax and two proper motions.  The uncertainties in all three affect the
inferred velocity and its uncertainty. Moreover, as the \gaia parallaxes and proper motions
generally have non-zero correlations, these must also be taken into account. This can be done in a
straightforward manner in the Bayesian approach, as is shown in the tutorial at
\subrepo{3d-distance}. This sets up a three-parameter model (distance, speed, angle) for a
source. Using the three measurements (parallax, two proper motions) in a multivariate Gaussian
likelihood, and suitable priors on the parameters, we can compute the trivariate posterior. This is
sampled in the posterior using an MCMC algorithm for a set of stars.

\subsection{Luminosity calibration}
\label{subsec:lumin-cal}

In this tutorial (\subrepo{luminosity-calibration}) the problem of inferring (or calibrating) the mean
absolute magnitude of a specific class of stars is treated. The measurements at hand are the
parallax and apparent magnitude for each of the stars in the sample and the task is to infer their
mean absolute magnitude $\mu_M$ and the spread $\sigma_M$ around this mean. This is very similar to
the problem that \cite{LutzKelker73} and \cite{1977A&A....56..273T} treated, and a Bayesian approach
to solving this problem was presented by \cite{2012asas.book.....V} (albeit with the use of improper
priors, which we  again note is bad practice). A more complex version of this problem (accounting
for extinction and a contaminating population of stars) and its Bayesian solution was presented in
\cite{2017MNRAS.471..722H}.  In this tutorial three important points are illustrated:
\begin{itemize}
  \item Often the explicit computation of the distances to stars is not of interest. In this example
    only the mean absolute magnitude of the stars is to be estimated, and the forward modelling
    approach as part of the Bayesian inference  avoids the need to calculate or estimate
    distances.
  \item The data for all the stars carry information on the mean absolute magnitude, including the
    negative parallaxes or parallaxes with large relative errors. This information can naturally be
    incorporated in a forward modelling approach (in this example as part of a Bayesian inference
    method), thus avoiding the introduction of truncation biases caused by the selection of stars
    with `good' parallaxes.
  \item If the selection function is known (in this example the survey is magnitude limited), it can
    and should be included in the forward modelling. This accounts for
    sample selection biases that would otherwise occur.
\end{itemize}

\subsection{Period-luminosity relation}
\label{subsec:PLrelation}

In this tutorial (\subrepo{period-luminosity-relation}) we include a hierarchical model to infer
period-luminosity-metallicity relations for classical pulsating stars. The full model can be applied
to fundamental mode RR~Lyrae stars and the abridged version (without the metallicity dependence) is
suitable for samples of classical Cepheids. We include the data set for the RR~Lyrae stars described
and used for inference in \cite{2017A&A...605A..79G} and \cite{HDP}. It contains a sample of 200
stars (including fundamental radial pulsators but also {\sl fundamentalised} first overtone
pulsators) with measured periods, apparent magnitudes in the $K$-band, metallicities, and parallaxes
from the TGAS catalogue. In the tutorial, we describe the hierarchical model and discuss potential
biases in the data set. Finally, we analyse the sensitivity of the results to different choices of
the priors and related parameters.

%----------------------------------------------------------------
% Conclusions and recommendations
%----------------------------------------------------------------
\section{Conclusions}
  \label{sec:conclusions}
  
\gaia data releases will provide a huge increase of astrometric data available for the scientific community. More than a billion parallaxes and proper motions allow new openings into many astronomical topics. In most cases astronomers are exploiting the \gaia catalogues to obtain physical quantities such as distance and velocity. Although it is easy to extract from the \gaia data, it is well known that direct inversion of parallax will lead to biases, which become more and more significant the larger the relative parallax uncertainty. While \gaia will provide high-quality astrometric measurements, hundreds of millions of stars have precisions which require proper statistical treatment in order to avoid biased conclusions. The aim of this paper is to guide the users of \gaia data to handle astrometric data correctly.

In this study we summarise methods used to avoid biases when converting astrometric data into physical quantities. Starting from simple, non-recommended, sample truncation to more complex methods, the biases associated with the methods are presented. The basic recommendation is to treat derivation of physical quantities from astrometric measurements as an inference problem, which should be preferably handled with Bayesian approach. The recommended methods are described in \secref{sec:guide} with a summary in \secref{sec:summary_recommendations}. To aid the users further, \secref{sec:tutorial} contains practical examples with links to Python and R code.

\gaia will provide fundamental data for many fields of astronomy. Further data releases will provide more data, and more precise data. Nevertheless, for full use of the potential it will always be necessary to pay careful attention to the statistical treatment of parallaxes and proper motions. The purpose of this paper is to help astronomers find the correct approach.

%----------------------------------------------------------------
% Acknowledgements
%----------------------------------------------------------------
\begin{acknowledgements}

This work has made use of results from the European Space Agency (ESA)
space mission {\it Gaia}, the data from which were processed by the {\it Gaia
Data Processing and Analysis Consortium} (DPAC).  Funding for the DPAC
has been provided by national institutions, in particular the
institutions participating in the {\it Gaia} Multilateral Agreement. The
{\it Gaia} mission website is \url{http: //www.cosmos.esa.int/gaia}. The
authors are current or past members of the ESA {\it Gaia} mission team and
of the {\it Gaia} DPAC.

This work was supported by the MINECO (Spanish Ministry of Economy) through 
grant ESP2016-80079-C2-1-R (MINECO/FEDER, UE) and ESP2014-55996-C2-1-R 
(MINECO/FEDER, UE) and MDM-2014-0369 of ICCUB (Unidad de Excelencia
`Mar\'{\i}a de Maeztu') and by the DLR (German space agency) via grant
50\,QG\,1403.

\end{acknowledgements}

%----------------------------------------------------------------
% Bibliography
%----------------------------------------------------------------
\bibliographystyle{aa} % style aa.bst
\bibliography{bibliography}

\begin{thebibliography}{38}
\expandafter\ifx\csname natexlab\endcsname\relax\def\natexlab#1{#1}\fi
\expandafter\ifx\csname url\endcsname\relax
  \def\url#1{{\tt #1}}\fi
\expandafter\ifx\csname urlprefix\endcsname\relax\def\urlprefix{URL }\fi

\bibitem[{{Anderson} et~al.(2017){Anderson}, {Hogg}, {Leistedt},
  {Price-Whelan}, \& {Bovy}}]{2017arXiv170605055A}
{Anderson} L., {Hogg} D.W., {Leistedt} B., {Price-Whelan} A.M., {Bovy} J., Jun.
  2017, ArXiv e-prints

\bibitem[{{Arenou} \& {Luri}(1999)}]{ABL}
{Arenou} F., {Luri} X., 1999, In: {Egret} D., {Heck} A. (eds.) Harmonizing
  Cosmic Distance Scales in a Post-HIPPARCOS Era, vol. 167 of Astronomical
  Society of the Pacific Conference Series, 13--32

\bibitem[{{Arenou} et~al.(2017){Arenou}, {Luri}, {Babusiaux}
  et~al.}]{Arenou_2017}
{Arenou} F., {Luri} X., {Babusiaux} C., et~al., Feb. 2017, \aap, 599, A50

\bibitem[{{Astraatmadja} \&
  {Bailer-Jones}(2016{\natexlab{a}})}]{2016ApJ...832..137A}
{Astraatmadja} T.L., {Bailer-Jones} C.A.L., Dec. 2016{\natexlab{a}}, \apj, 832,
  137

\bibitem[{{Astraatmadja} \&
  {Bailer-Jones}(2016{\natexlab{b}})}]{2016ApJ...833..119A}
{Astraatmadja} T.L., {Bailer-Jones} C.A.L., Dec. 2016{\natexlab{b}}, \apj, 833,
  119

\bibitem[{{Bailer-Jones}(2015)}]{2015PASP..127..994B}
{Bailer-Jones} C.A.L., Oct. 2015, \pasp, 127, 994

\bibitem[{{Bovy}(2017)}]{2017MNRAS.470.1360B}
{Bovy} J., Sep. 2017, \mnras, 470, 1360

\bibitem[{{Brown}(2012)}]{2012asas.book.....V}
{Brown} A.G.A., Nov. 2012, {Statistical Astrometry}

\bibitem[{{Brown} et~al.(1997){Brown}, {Arenou}, {van Leeuwen}, {Lindegren}, \&
  {Luri}}]{1997ESASP.402...63B}
{Brown} A.G.A., {Arenou} F., {van Leeuwen} F., {Lindegren} L., {Luri} X., Aug.
  1997, In: {Bonnet} R.M., {H{\o}g} E., {Bernacca} P.L., et~al. (eds.)
  Hipparcos - Venice '97, vol. 402 of ESA Special Publication, 63--68

\bibitem[{{de Bruijne} et~al.(2001){de Bruijne}, {Hoogerwerf}, \& {de
  Zeeuw}}]{2001A&A...367..111D}
{de Bruijne} J.H.J., {Hoogerwerf} R., {de Zeeuw} P.T., Feb. 2001, \aap, 367,
  111

\bibitem[{{Delgado} et~al.(2018){Delgado}, {Sarro}, {Clementini}, {Muraveva},
  \& {Garofalo}}]{HDP}
{Delgado} H.E., {Sarro} L.M., {Clementini} G., {Muraveva} T., {Garofalo} A.,
  Mar. 2018, ArXiv e-prints

\bibitem[{{Eddington}(1913)}]{1913MNRAS..73..359E}
{Eddington} A.S., Mar. 1913, \mnras, 73, 359

\bibitem[{{ESA}(1997)}]{1997ESASP1200.....E}
{ESA} (ed.), 1997, {The HIPPARCOS and TYCHO catalogues. Astrometric and
  photometric star catalogues derived from the ESA HIPPARCOS Space Astrometry
  Mission}, vol. 1200 of ESA Special Publication (available online at
  \url{https://www.cosmos.esa.int/hipparcos})

\bibitem[{{Gaia Collaboration} et~al.(2016){Gaia Collaboration}, {Brown},
  {Vallenari} et~al.}]{2016A&A...595A...2G}
{Gaia Collaboration}, {Brown} A.G.A., {Vallenari} A., et~al., Nov. 2016, \aap,
  595, A2

\bibitem[{{Gaia Collaboration} et~al.(2017{\natexlab{a}}){Gaia Collaboration},
  {Clementini}, {Eyer} et~al.}]{2017A&A...605A..79G}
{Gaia Collaboration}, {Clementini} G., {Eyer} L., et~al., Sep.
  2017{\natexlab{a}}, \aap, 605, A79

\bibitem[{{Gaia Collaboration} et~al.(2017{\natexlab{b}}){Gaia Collaboration},
  {van Leeuwen}, {Vallenari} et~al.}]{2017A&A...601A..19G}
{Gaia Collaboration}, {van Leeuwen} F., {Vallenari} A., et~al., May
  2017{\natexlab{b}}, \aap, 601, A19

\bibitem[{{Gaia Collaboration} et~al.(2018{\natexlab{a}}){Gaia Collaboration},
  {Arenou}, {Luri} et~al.}]{DR2-DPACP-39}
{Gaia Collaboration}, {Arenou} F., {Luri} X., et~al., Apr. 2018{\natexlab{a}},
  \aap\ (special issue for Gaia DR2)

\bibitem[{{Gaia Collaboration} et~al.(2018{\natexlab{b}}){Gaia Collaboration},
  {Brown}, {Vallenari}, {Prusti}, \& {et al.}}]{DR2-DPACP-36}
{Gaia Collaboration}, {Brown} A.G.A., {Vallenari} A., {Prusti} T., {et al.},
  Apr. 2018{\natexlab{b}}, \aap\ (special issue for Gaia DR2)

\bibitem[{{Gaia Collaboration} et~al.(2018{\natexlab{c}}){Gaia Collaboration},
  {Lindegren}, {Hern{\'a}ndez} et~al.}]{DR2-DPACP-51}
{Gaia Collaboration}, {Lindegren} L., {Hern{\'a}ndez} J., et~al., Apr.
  2018{\natexlab{c}}, \aap\ (special issue for Gaia DR2)

\bibitem[{{Hawkins} et~al.(2017){Hawkins}, {Leistedt}, {Bovy}, \&
  {Hogg}}]{2017MNRAS.471..722H}
{Hawkins} K., {Leistedt} B., {Bovy} J., {Hogg} D.W., Oct. 2017, \mnras, 471,
  722

\bibitem[{{Iorio} et~al.(2018){Iorio}, {Belokurov}, {Erkal}
  et~al.}]{2018MNRAS.474.2142I}
{Iorio} G., {Belokurov} V., {Erkal} D., et~al., Feb. 2018, \mnras, 474, 2142

\bibitem[{{Leistedt} \& {Hogg}(2017{\natexlab{a}})}]{2017AJ....154..222L}
{Leistedt} B., {Hogg} D.W., Dec. 2017{\natexlab{a}}, \aj, 154, 222

\bibitem[{{Leistedt} \& {Hogg}(2017{\natexlab{b}})}]{2017arXiv170308112L}
{Leistedt} B., {Hogg} D.W., Mar. 2017{\natexlab{b}}, ArXiv e-prints

\bibitem[{{Lindegren} et~al.(2000){Lindegren}, {Madsen}, \&
  {Dravins}}]{2000A&A...356.1119L}
{Lindegren} L., {Madsen} S., {Dravins} D., Apr. 2000, \aap, 356, 1119

\bibitem[{{Lindegren} et~al.(2012){Lindegren}, {Lammers}, {Hobbs}
  et~al.}]{2012A&A...538A..78L}
{Lindegren} L., {Lammers} U., {Hobbs} D., et~al., Feb. 2012, \aap, 538, A78

\bibitem[{{Lindegren} et~al.(2016){Lindegren}, {Lammers}, {Bastian}
  et~al.}]{2016A&A...595A...4L}
{Lindegren} L., {Lammers} U., {Bastian} U., et~al., Nov. 2016, \aap, 595, A4

\bibitem[{{Lutz} \& {Kelker}(1973)}]{LutzKelker73}
{Lutz} T.E., {Kelker} D.H., Oct. 1973, \pasp, 85, 573

\bibitem[{{Malmquist}(1920)}]{1920MeLuS..22....3M}
{Malmquist} G.K., Mar. 1920, Meddelanden fran Lunds Astronomiska Observatorium
  Serie II, 22, 3

\bibitem[{{Palmer} et~al.(2014){Palmer}, {Arenou}, {Luri}, \&
  {Masana}}]{2014A&A...564A..49P}
{Palmer} M., {Arenou} F., {Luri} X., {Masana} E., Apr. 2014, \aap, 564, A49

\bibitem[{{Robin, C.} et~al.(2012){Robin, C.}, {Luri, X.}, {Reyl{\'e}, C.}
  et~al.}]{ACRXLCR-2012}
{Robin, C.}, {Luri, X.}, {Reyl{\'e}, C.}, et~al., 2012, A{\&}A, 543, A100

\bibitem[{{Sch{\"o}nrich} \& {Aumer}(2017)}]{2017MNRAS.472.3979S}
{Sch{\"o}nrich} R., {Aumer} M., Dec. 2017, \mnras, 472, 3979

\bibitem[{{Sch{\"o}nrich} et~al.(2012){Sch{\"o}nrich}, {Binney}, \&
  {Asplund}}]{2012MNRAS.420.1281S}
{Sch{\"o}nrich} R., {Binney} J., {Asplund} M., Feb. 2012, \mnras, 420, 1281

\bibitem[{{Schr{\"o}der} et~al.(2004){Schr{\"o}der}, {Kaper}, {Lamers}, \&
  {Brown}}]{2004A&A...428..149S}
{Schr{\"o}der} S.E., {Kaper} L., {Lamers} H.J.G.L.M., {Brown} A.G.A., Dec.
  2004, \aap, 428, 149

\bibitem[{{Sesar} et~al.(2017){Sesar}, {Fouesneau}, {Price-Whelan}
  et~al.}]{2017ApJ...838..107S}
{Sesar} B., {Fouesneau} M., {Price-Whelan} A.M., et~al., Apr. 2017, \apj, 838,
  107

\bibitem[{{Smith} \& {Eichhorn}(1996)}]{SmithEichhorn}
{Smith} H. Jr., {Eichhorn} H., Jul. 1996, \mnras, 281, 211

\bibitem[{{Trumpler} \& {Weaver}(1953)}]{1953stas.book.....T}
{Trumpler} R.J., {Weaver} H.F., 1953, {Statistical astronomy}

\bibitem[{{Turon Lacarrieu} \& {Cr\'ez\'e}(1977)}]{1977A&A....56..273T}
{Turon Lacarrieu} C., {Cr\'ez\'e} M., Apr. 1977, \aap, 56, 273

\bibitem[{{Zinn} et~al.(2017){Zinn}, {Huber}, {Pinsonneault}, \&
  {Stello}}]{2017ApJ...844..166Z}
{Zinn} J.C., {Huber} D., {Pinsonneault} M.H., {Stello} D., Aug. 2017, \apj,
  844, 166

\end{thebibliography}

%------------
% Appendices
%------------
\begin{appendix}

\section{Description of the simulated samples used in this paper}
  \label{sec:app_samples}
        
The data used in this paper is from a {\it Gaia Universe Model Snapshot} (GUMS) simulation \citep{ACRXLCR-2012}, together with the {\it Gaia}-like uncertainties
and an estimation of the observable data. The uncertainties were computed from an implementation of the recipes described in the {\it Gaia} Science performance web page\footnote{https://www.cosmos.esa.int/web/gaia/science-performance}, provided by the python PyGaia toolkit\footnote{https://pypi.python.org/pypi/PyGaia/}, which then  have to be re-scaled to fit the \gdrtwo expectations.

\begin{figure}[htb]
        \centering
                \includegraphics[width = 1.00\columnwidth]{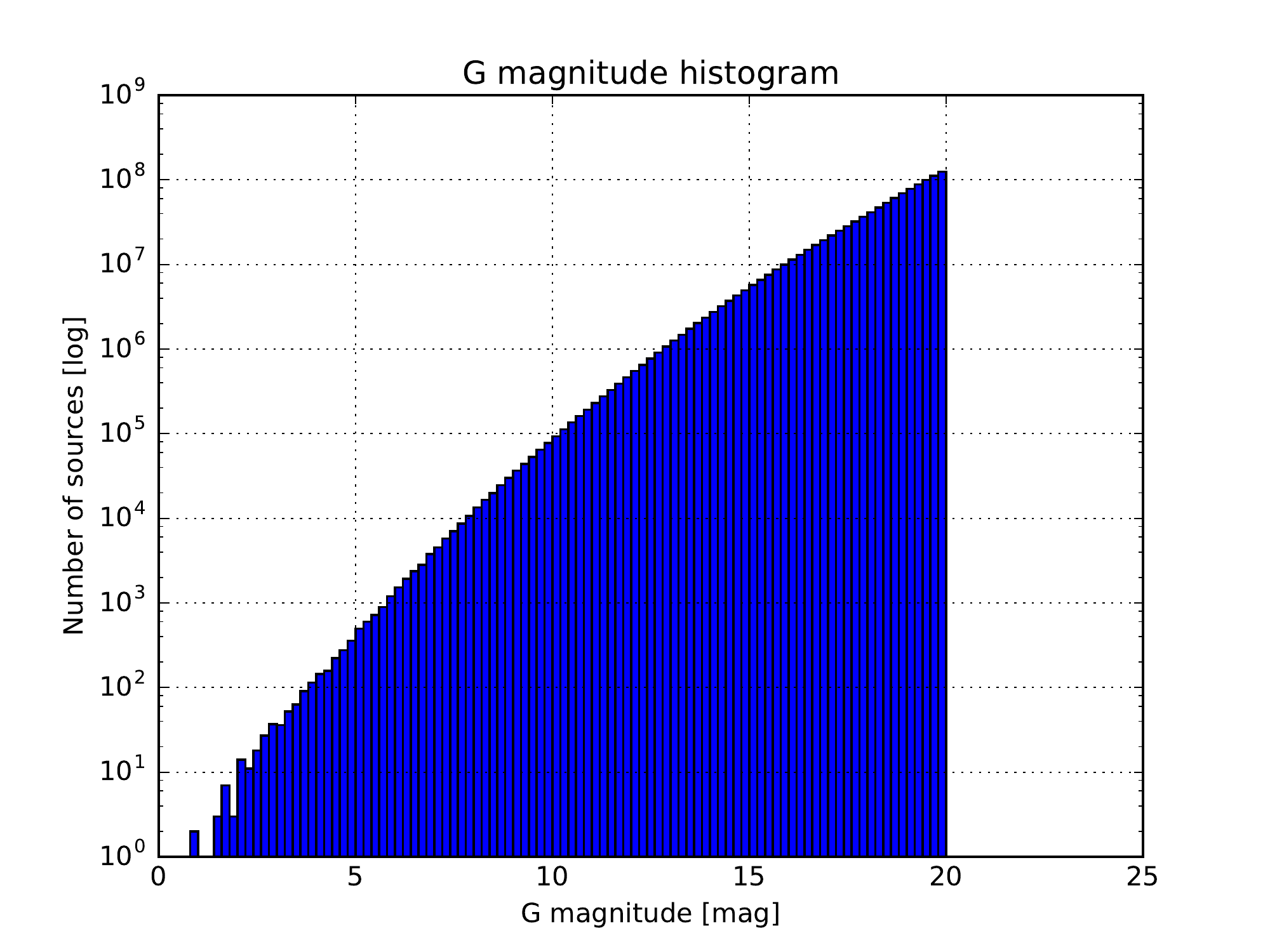}
                \caption{Histogram of the stars' $G$ magnitude. The total number of sources is $1\,069\,138\,714$, which are distributed in bins of size $\Delta G = 0.2$; the sample is limited to $G < 20$.}
                \label{fig:g_hist}
\end{figure}

%simulation contents + plots (magnitude)
The simulation contains around $10^9$ sources including only single stars, {i.e.} stars not belonging to multiple systems, up to $G < 20$ magnitude, distributed as shown in  \figref{fig:g_hist}.

%error description + plots (error as function of magnitude DR2)
To compute the {\it Gaia} potential observables, standard uncertainties must be added to the simulation. Because astrometric quantities (positions, parallaxes, and proper motions) are related, a single formalism to derive its standard errors is needed. PyGaia implements a simple performance model depending on the $V\text{--}I_C$ colour term and the $G$ magnitude to estimate the end-of-mission errors for the parallax uncertainty:
\begin{equation}
        \sigma_{\varpi} \text{[$\mu$as]} = \sqrt{-1.631 + 680.766 \, z + 32.732 \, z^2} \, [0.986 +\left(1-0.986\right)\left(V\text{-}I_C\right)],
        \label{eq:sigmapi}
\end{equation}
with
\begin{equation}
        z = \text{max(}10^{0.4\left(12.09 - 15\right)}\text{,}10^{0.4\left(G-15\right)}\text{)}
        \label{eq:z}
\end{equation}
(see the {\it Gaia} science performance web for more details). It also takes into account the variation of the uncertainties over the sky because of the scanning law, tabulated as a function of the ecliptic latitude\footnote{https://www.cosmos.esa.int/web/gaia/table-2-with-ascii} $\beta$. 

\begin{figure}[htb]
        \centering
                \includegraphics[width = 1.00\columnwidth]{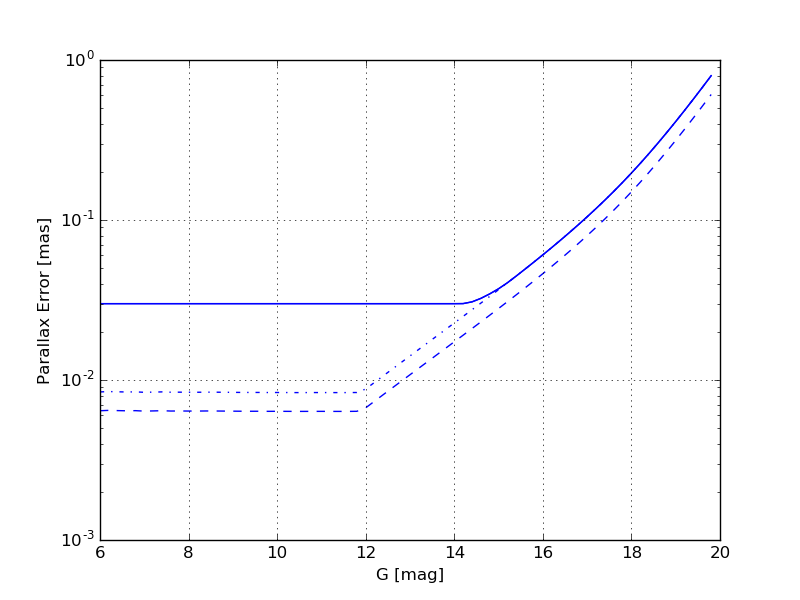}
                \caption{Average estimated errors as function of $G$. The dash-dotted line represents the uncertainty at \gdrtwo, while the dashed line represents the end-of-mission uncertainties. The solid line represents \gdrtwo errors, including systematics.}
                \label{fig:errors_g}
\end{figure}

However, these are end-of-mission uncertainties, so they have to be scaled by the fraction of mission time completed in order to get an estimation of them for \gdrtwo. In the case of the parallax, only the factor $\sqrt{\frac{5}{L}}$, being $L$ the mission time (years) included in \gdrtwo, need to be applied; this error model is further described in \cite{Arenou_2017}. In our case we have also updated the calibration floor  to take into account the properties of the \gdrtwo formal errors, as shown in \figref{fig:errors_g}. This calibration floor introduces a minimum formal error stemming from the fact that 
the calibrations used in the data processing (models and parameters) are at this stage still being refined. This floor affects mainly bright stars, while for faint stars the photon noise dominates. \figref{fig:errors_g} shows the model used in the simulation for the \gdrtwo parallax uncertainties as a function of the $G$ magnitude.

\end{appendix}

\end{document}